# Influence of Entropy Changes on First Passage Time in the Thermodynamics of Trajectories

V. V. Ryazanov

Institute for Nuclear Research, pr. Nauki, 47, Kiev, Ukraine, e-mail: vryazan19@gmail.com

An ensemble of trajectories with dynamical activity and first-passage time (*FPT*) is considered in the context of the thermodynamics of trajectories. The relationship between the average *FPT* and the total change in entropy is determined, with a focus on non-negative values of the total change in entropy only. Similar dependencies were found for dynamic activity, the dispersion of dynamic activity, and *FPT*, as well as for the correlation between dynamic activity and *FPT*. Applying the obtained results to model systems reveals dependencies on entropy changes in stationary nonequilibrium and equilibrium states. To relate changes in entropy to the conjugate parameter of the *FPT*, three models of the distribution function are applied to classical two- and three-level systems, and to a quantum two-level system.

## 1. Introduction

In the description of nonequilibrium phenomena, the so-called "thermodynamics of trajectories" approach is used [1-28]. In this approach, the behavior of time-ordered dynamic events is characterized similarly to the thermodynamic description of configurations in space. In a non-equilibrium situation, an analog of equilibrium thermodynamic quantities and relations is used, but for the dynamic case, trajectories - rather than states or configurations - become the object of consideration. The role of volume in this case is played by time.

Using the methods of large deviations (*LD*) [28], ensembles of trajectories can be classified based on dynamic order parameters and their conjugate fields. This is Ruelle's thermodynamic formalism [27], applied to the space of trajectories, not configurations. Quantities similar to the free energy density and entropy density have been identified and used to understand the rare dynamical behavior of systems, both classical [7–21] and quantum [22–26]. This approach has effectively described the dynamics of glassy systems [14, 29–31]. Statistical ensembles with fluctuating times of the deployment of trajectories were considered in [32]. *FPT* for dynamic ensembles of trajectories is considered in [33].

In [33], focus on trajectory observables defined in terms of the jumps in a trajectory, this is called counting observables. A more general definition of fluctuations of observables of the trajectory *X* is given in [34]. Dissipation bounds and current fluctuations are considered in [35-41]. An example is the total number of changes in the trajectory configuration or dynamic activity [8, 2, 12], sometimes also called "traffic" [42-44]. In the thermodynamics of trajectories, an *s*-ensemble is singled out, in which the random variable is a time-additive value *A*, for example, dynamic activity *K*, and the time *τ* of the deployment of trajectories is fixed, and an *x*-ensemble, in which the values *A* or *K* are fixed, and the random variable is time *τ*. In [33], *τ* is considered as the *FPT*. In this article, this approach is used. In *x*-ensemble, the random time *τ* is *FPT* when a fixed value of *K* is reached. In [32] a generalization of the *x*-ensemble was obtained for multiple observables. In the present article, one more generalization of such an ensemble is carried out.

In papers [83–84, 52–53], *FPT* is considered as a thermodynamic variable and included in the generalized Gibbs distribution. Similar results are obtained in [33, 9, 88] for the joint distribution of the activity parameters and the process time. The characteristics of *FPT* are affected by various factors. One of the main factors is the entropy change, which includes both intrasystem entropy changes and entropy exchange with the environment. This paper studies the effect of entropy changes on the behavior of nonequilibrium systems. The entropy change during *FPT* is taken into account. Three partition function models are used. The dependencies of the mean *FPT*, *FPT* variance, *FPT* correlation, and dynamic activity on the total entropy change are calculated. These dependencies depend on the interaction with the environment and change with a change in the sign of the entropy flows describing the exchange between the system and the environment.

Typically, average *FPT* values are calculated from the Laplace transform of the *FPT* probability density by differentiating the Laplace transform and then setting the Laplace transform parameter to zero. In [83-84], the Laplace transform of the probability density *FPT* acts as the nonequilibrium part of the partition function, and the Laplace transform parameter is not equal to zero. The main idea of this article is that the parameter $x$ (in notation [2, 32]) or $\gamma=x$ (in notation [83-84]) which conjugate to the random variable time $\tau$ or *FPT* associated with changes in entropy and is equal to zero only in equilibrium, as is shown in [83-84]. The presence of a non-zero value of $\gamma$ corresponds to taking into account the change in entropy during the *FPT*. In works [85-86, 106], using examples of specific physical systems, it is shown that the influence of such consideration can be significant. Any nonequilibrium process, including the *FPT* process, occurs with a change in entropy, which is taken into account in this article. Just as the inverse temperature parameter $\beta=1/k_BT$ conjugate to the random energy is not equal to zero in the general case, so the parameter $\gamma$ (or $x$) is equal to zero only in the equilibrium case. Therefore, in the dependencies of the moments of random variables *FPT* (or $\tau$), of the dynamic activity $K$, as well as the parameters $M$ from (22)-(28) in the general case, non-zero values of $\gamma=x$ be taken into account.

The article introduces a statistical ensemble with fluctuating values of *FPT* and dynamic activity $K$ in the form (29) (also (39), (80)). With the help of this ensemble, explicit dependencies of the average values, variances, and correlations of random variables *FPT* and $K$ on the total change in entropy $\Delta s_{tot} = \Delta s_{sys} + \Delta s_m$ (38) during the *FPT* time (or $\tau$) are determined. The total change in entropy is related to the parameter $\gamma$. The parameter $\gamma$ is expressed through $\Delta s_{tot}$. The dependencies of the moments of random variables *FPT* and $K$ on the parameter $\gamma$ (or $x$) conjugate to *FPT* are expressed through the dependencies of these moments on the total change in entropy during the time *FPT* (or $\tau$). Expressions for the correlation between dynamic activity and *FPT* are obtained. It is physically obvious that such a correlation is not zero. Such dependencies of system parameters on the total change in entropy during *FPT* were previously unknown. The results obtained can be considered as an application of distributions from [83, 84] to the thermodynamics of trajectories. The application of this approach to specific physical problems was carried out, for example, in works [85-86, 106].

The article is organized as follows. Section 2 provides information on the thermodynamics of trajectories: *s*-ensembles and *x*-ensembles. In Section 3, a joint distribution of $K$ and *FPT* activities was obtained, from which the average values of *FPT* and $K$, their second moments, variances, and correlations between them were recorded. In the fourth section, dependencies of the first and second moments of distributions on the change in entropy are obtained, and curves are calculated for given numerical values of parameters. The fifth section contains brief conclusions.

## 2. Thermodynamics of Trajectories, *s*-ensembles and *x*-ensembles

This section provides an introduction to the key concepts and frameworks of thermodynamics of trajectories, with a focus on *s*-ensembles and *x*-ensembles. The *s*-ensemble provides the statistics of observable behavior over fixed time scales, enabling us to calculate average values, variances, and higher moments concerning the time duration $\tau$.

Key considerations in *s*-ensembles include:
- Observables are collected over a fixed time, enabling the determination of average rates and fluctuations typical for that timespan.
- Observables are assumed to exhibit typical distributions characterized by certain statistical properties as determined by the system's dynamics.

In contrast, an *x*-ensemble is defined by fixing the observable (such as a specific value of dynamic activity *K*) and allowing the time duration $\tau$ (or *FPT*) to vary. This approach acknowledges that the time it takes for a system to reach a specific dynamical state can vary, and understanding this variability is crucial for analyzing kinetics under non-equilibrium conditions.

In the context of the *x*-ensembles:
- The random variable (time $\tau$ or *FPT*) is treated as a crucial parameter that detects the fluctuations of the system in response to fixed observable values.
- This approach allows for calculations of mean first passage times, and their corresponding variances based on the predetermined observables.

A fundamental aspect of this framework is the joint distribution of the dynamic activity *K* and the *FPT*, which helps to establish correlations between these two observables. The study of such joint distributions sheds light on the thermodynamic behavior of the system as it relates to timescales and dynamic orders.

In this article, we explore how changes in entropy, associated with intrasystem interactions and environmental exchanges, impact the *FPT* and dynamic activity within both *s*- and *x*-ensembles. By linking these thermodynamic identities, we can derive expressions that reveal how entropy changes affect the overall kinetics of the system, including:
- Average *FPT* as a function of entropy change.
- Variance of *FPT* and its correlation with dynamic activity.
- How these relationships shift under different non-equilibrium conditions, highlighting the importance of the entropy flow's sign.

In summary, the thermodynamics of trajectories framework, through the precise definitions of the *s*-ensembles and *x*-ensembles, provides the means to analyze dynamic processes' statistical properties. These insights allow for a deeper understanding of how entropy changes inform the *FPT* and dynamic activity, linking microscopic dynamics to macroscopic thermodynamic principles. The following sections will delve into the mathematical derivations and empirical implications of these concepts.

The formalism of nonequilibrium thermodynamics along trajectories is analogous to the equilibrium thermodynamic formalism [27]. This article examines fluctuations in dynamic variables, such as dynamic activity in a glassy system, integrated over a long-time *t* and a large (but finite) system. In [2], the statistical properties of evolution and the dynamic history of the system are studied. Equilibrium thermodynamics considers the probability distribution over the configurations of a large system. In trajectory thermodynamics, the thermodynamic formalism is applied to probability distributions over trajectory motion histories.

In references [2, 9, 32, 49, 50], the classical stochastic system is described by the Master Equation:

$$\partial_t |P(t)\rangle = W |P(t)\rangle. \tag{1}$$

The vector $|P(t)\rangle$ represents the probability distribution at time $t$, $|P(t)\rangle = \sum_C P(C,t)|C\rangle$, where $P(C,t)$ describes the probability that the system is in configuration $C$ at time $t$, and $\{|C\rangle\}$ is the orthonormal basis of the configuration, $\langle C|C'\rangle = \delta_{CC'}$. For concreteness, we will focus on continuous-time Markov chains, but generalizations of what we describe below are simple. The master operator $W$ is the matrix:

$$W = \sum_{C' \neq C} w(C \to C')|C'\rangle\langle C| - \sum_C R(C)|C\rangle\langle C|. \tag{2}$$

Here $w(C \to C')$ is the rate of transition from $C$ to $C'$, and $R(C) = \sum_{C'} w(C \to C')$ is the rate of exit from $C$. In this description, the expectation of the operator $T$ is defined as $\langle T(t)\rangle = \langle -|T|P(t)\rangle$, where $\langle -| \equiv \sum_C \langle C|$ (such that $\langle -|P(t)\rangle = 1$ due to conservation of probability).

In the Boltzmann-Gibbs theory, the macroscopic properties of large systems are characterized by defining the statistical properties (mean and fluctuations) of extensive observables, such as energy or particle number. The microcanonical approach considers the properties of a system with a fixed total energy $E$. These properties are determined from the counting factor of the number of configurations with energy $E$ and the number of particles $N$, which represents the size (volume) of the system. In the limit of large sizes ($N \to \infty$), the entropy density $s(e) = \lim_{N \to \infty} (1/N) \ln \Omega(eN, N)$ is determined, which represents the relative weight of configurations with energy density $e$; $\Omega(E, N)$ is a number of configurations with energy $E$, where $N$ represents the size (the volume) of the system. In a dynamic context, the history of the system is considered between the initial time $\tau = 0$ and the end time $\tau$. Instead of considering the statistics of the energy $E$, we study the observed $A$, which is extensive over the observation time $\tau$. The dynamic analog of the quantity $\Omega(E, N)$ is the probability distribution of this observable, where the value $\Omega_{dyn}(A, \tau)$ represents the fraction of histories with a given value of the time-extensive observable $A$.

At the mathematical level, the choice of the observable $A$ is somewhat arbitrary, but the application of the thermodynamic formalism requires that the quantity $\log \Omega_{dyn}(A = a\tau, \tau)/\tau$ have a finite limit for large times $\tau$. Given this limitation, the choice of the order parameter $A$ is determined by physical considerations: we must use an observable that reveals the main physical processes occurring in the system. For example, in non-equilibrium systems in contact with two reservoirs of particles, we can define $A$ as the total particle flux, which is the number of particles transferred from one reservoir to the other between times 0 and $\tau$. In the context of glassy phenomena, we examine observables that measure the activity or complexity of the system's history [7, 8, 12].

In the Boltzmann-Gibbs approach, the canonical ensemble is defined in terms of the partition function:

$$Z(\beta, N) = \sum_E \Omega(E, N) e^{-\beta E}, \tag{3}$$

which characterizes the system at a given temperature $\beta^{-1}$. Phase transitions depend on intensive free energy $f(\beta) = -\lim_{N \to \infty} \ln Z(\beta, N)/\beta N$ singularities. The dynamical analog of this thermodynamic partition function (3) is:

$$Z_A(s, \tau) = \sum_A \Omega_{dyn}(A, \tau) e^{-sA}, \tag{4}$$

where we introduce an intensive field *s* conjugate to *A*. This field plays a role analogous to the inverse temperature *β*. The dynamic partition function $Z_A(s,\tau)$ is a key concept in Ruelle's thermodynamic formalism [27]. A correspondence is established between the thermodynamic limit of a large system size ($N\to\infty$) and the limit of a long time ($\tau\to\infty$) in Ruelle's formalism. In [2], systems are considered where a large time limit is taken at a fixed system size, and in some cases, a second limit of the large system size *N* is also taken. If we consider systems without thermodynamic phase transitions, then no singular behavior occurs when taking the limit of large *N* at a fixed *τ*. In this case, we expect the limits of large *N* and large *τ* to commute, but in general, this is not the case. In this article, we first find the average values and then take the limit of large *τ*.

The duration of the system's history is characterized by its length *τ*. The probability of measuring the value *A* for an observable *A* in a history of length *τ* is given by:

$$\Omega_{dyn}(A,\tau) = \sum_C P(C,A,\tau). \tag{5}$$

The value $P(C,A,\tau)$ (generalizing the probability $P(C,\tau)$) is defined as the probability of being in configuration *C* at time *τ*, having measured a value *A* of the time-extensive variable between 0 and *τ*.

The sequence of configurations $C_0 \to \ldots \to C_K$ and the sequence of jump times $t_1,\ldots,t_K$ determine the system's history with duration *τ*. The time record of configurations and waiting times of jumps between them observed up to time *τ*, describes the trajectory of total time *τ*. If such a trajectory is denoted by $X_\tau$, then $X_\tau=(C_0 \to C_{t_1} \to \ldots C_{t_n})$, where $C_0$ is the initial configuration, and $t_i$ is the time of the transition from $C_{t_{i-1}}$ to $C_{t_i}$ (so that the waiting time of the *i*-th jump is $t_i - t_{i-1}$). The trajectory $X_\tau$ changes configurations only *n* times, and $t_n \leq \tau$ there is no jump between $t_n$ and *τ*. Between the times $t_k$ and $t_{k+1}$, the system remains in its configuration $C_k$.

If we consider the statistics of time-extensive observables [7, 8], we can study the properties of dynamics. Dynamic activity *K*, defined as the total number of configuration changes per trajectory [8, 2, 12], is one such observable. In [34], the fluctuations of observables of trajectory *X* are written in the form:

$$A(t) = \sum_{x\neq y} a_{xy} N_{xy}(t), \tag{6}$$

where $a_{xy}$ are arbitrary real numbers with $\sum_{x\neq y} a_{xy} > 0$, and $N_{xy}(t)$ represents the elementary fluxes, which is the number of jumps from *x* to *y* up to time *τ* in $X_\tau$. For time-integrated currents, the coefficients are antisymmetric, while for counting observables (such as the activity), they are symmetric. The value of *K* is determined by equation (6) as $a_{xy}=1$. Below, we will consider a special case of arbitrary values *A* and dynamic activity *K*. We can generalize this to the general case. The distribution of *K* over all trajectories $X_\tau$ of total time *τ* is given by:

$$P_\tau(K) = \sum_{X_\tau} \delta(K - \hat{K}[X_\tau])P(X_\tau), \tag{7}$$

where the probability $P(\mathbf{X}_\tau)$ is the probability of observing this trajectory out of all the possible ones of total time *τ*.

The operator $\hat{K}$ counts the number of jumps in a trajectory, which is the number of times the system changes configuration. The distribution $P_\tau(K)=\Omega_{dyn}(K,\tau)$ is given by (4)-(5). For large *τ* this probability takes on a large deviations form [8, 21, 28], given by:

$$P_\tau(K) \sim e^{-\tau\varphi(K/\tau)}. \tag{8}$$

Equivalent information is contained in the generating function (4),
$$Z_\tau(s) \equiv \sum_K e^{-sK} P_\tau(K) = \sum_{X_\tau} e^{-s\hat{K}[X_\tau]} P(X_\tau), \quad (9)$$

whose derivatives give the moments of the activity, $\langle K^n \rangle = (-1)^n \partial_s^n Z_t(s)|_{s=0}$. For large $\tau$, the generating function also takes on a *LD* form [28, 8], given by:
$$Z_\tau(s) \sim e^{\tau \theta(s)}. \quad (10)$$

In the large time limit, the behavior of $\hat{P}_A(C,s,t) = \sum_A e^{-sA} P(C,A,t)$ is assumed to be $\hat{P}_A(C,s,\tau) \sim R_0(C,s) e^{\tau \theta_A(s)}$ where $\theta_A(s)$ is the largest eigenvalue of $W_A$ [2] and $R_0(C,s)$ is the associated right eigenvector. Thus, for large times,
$$Z_A(s,\tau) = \sum_C \hat{P}(C,A,\tau) \sim e^{\tau \theta_A(s)}, \quad (11)$$

the function $\theta_A(s)$ is considered to be a (negative) dynamic free energy per unit time, which measures the system's energy per unit time. Probability conservation implies that $Z_A(0,\tau) = 1$, so $\theta(0) = 0$ for all stochastic systems, which means that the probability of the system being in any configuration is conserved over time.

From (10), the scaled cumulant generation function for the activity is determined, which gives the *n*-th activity cumulant (per unit time):
$$\frac{\langle\langle K^n \rangle\rangle}{\tau} = (-1)^n \frac{\partial^n}{\partial s^n} \theta(s)|_{s=0}, \quad (12)$$

where $\langle\langle . \rangle\rangle$ indicates cumulant (mean, variance, etc.). This contains the full statistical information about *K*, which is a measure of the system's activity.

Entropy and free energy in the Boltzmann-Gibbs theory are related through the Legendre transformation (as can be seen from (3)). The function $\theta(s)$ from (10) is a dynamical analog of the free energy density $f(\beta)$. Just like thermodynamic potentials, the *LD* functions $\varphi(k)$ and $\theta(s)$ are related by a Legendre-Fenchel transformation [8, 27], given by:
$$\varphi(k) = -\min_s (\theta(s) + sk) \quad (13)$$

together with the inversion formula:
$$\theta(s) = -\min_k (\varphi(k) + sk). \quad (14)$$

The above applies to the so-called *s*-ensembles, where the time $\tau$ of the system's history with duration $\tau$ is fixed [2, 32]. In [32], *x*-ensembles are introduced, where *K* is fixed and the time $\tau$ fluctuates. If we denote by $Y_K$ a trajectory $Y_K = (C_0 \to C_{t_1} \to ... \to C_K)$, where the number of configuration changes is fixed to be *K*, but the time $\tau$ of the final *K*-th jump fluctuates from one trajectory to another. From (1)-(2), the probability of $Y_K$ is given by:
$$P(Y_K) = p_0(C_0) \prod_{i=1}^K e^{-(t_i - t_{i-1}) R(C_{t_{i-1}})} w(C_{t_{i-1}} \to C_{t_i}), \quad (15)$$

where $t_0=0$ and $t_K = \tau$. The distribution $P_K(\tau)$ of total trajectory length $\tau$ for fixed activity *K* is given by:

$$P_K(\tau) = \sum_{Y_\tau} \delta(\tau - \hat{\tau}[Y_K]) P(Y_K) =$$
$$= \sum_{C_0 \ldots C_K} p_0(C_0) \prod_{i=1}^{K-1} \int_0^{t_{i+1}} dt_i\, e^{-(t_i - t_{i-1}) R(C_{t_{i-1}})} w(C_{t_{i-1}} \to C_{t_i}). \tag{16}$$

For large $K$, this probability takes on a *LD* form,
$$P_K(\tau) \sim e^{-K\Phi(\tau/K)}. \tag{17}$$

The corresponding moment generating function for $\tau$ is given by:
$$Z_K(x) \equiv \int_0^\infty d\tau\, e^{-x\tau} P_K(\tau)$$
$$= \sum_{Y_K} e^{-x\hat{\tau}[Y_K]} P(Y_K), \tag{18}$$

$\langle \tau^n \rangle = (-1)^n \partial_x^n Z_K(x)|_{x=0}$. For large $K$, the generating function also takes on a *LD* form,
$$Z_K(x) \sim e^{Kg(x)}. \tag{19}$$

Equation (18) is the partition sum for the ensemble of trajectories with probabilities, which is a measure of the probability of each trajectory:
$$P_x(Y_K) \equiv Z_K^{-1}(x) e^{-x\hat{\tau}[Y_K]} P(Y_K). \tag{20}$$

The function $g$ is the functional inverse of $\theta$ and vice versa [32], given by:
$$\theta(s) = g^{-1}(s), \quad g(\gamma) = \theta^{-1}(\gamma),$$
$$s = g(\gamma), \quad \gamma = \theta(s). \tag{21}$$

We replace $g(x)$ with $g(\gamma)$ (to use the notation $\gamma$ from [83-84] for the parameter conjugate to *FPT*).

## 3. Joint Distribution for Activity and FPT

Above we considered *s*-ensembles and *x*-ensembles. This section, following [32], examines the generalization of the *x*-ensemble to multiple observables. Next, a joint distribution was derived for the random variables representing dynamic activity $K$ and fluctuating total time. This distribution is derived using the conditional probability formula and the effective parameter $s_{ef}$.

Now, consider the statistics of first-passage times (*FPT*) (also called stopping times) [45-84]. This is the time when a certain observed trajectory first reaches the threshold value. In the thermodynamics of trajectories, in this situation, ensembles of trajectories with total fixed time are replaced by ensembles of trajectories of fluctuating total time.

In [32], the case is considered of the statistics of several different time-extensive quantities [2, 12]. For example, one could think of counting, instead of the total activity, the total number of certain kinds of transitions, or the time integral of a certain quantity such as the energy. Suppose there are $N$ different dynamical observables, which we denote collectively by the vector $\vec{M} \equiv (M_1, \ldots, M_N)$. Under the dynamics Eqs. (1)-(2) there will be a joint probability for observing a combination of these $\vec{M} \equiv (M_1, \ldots, M_N)$ quantities, $P_\tau(\vec{M})$. For large $\tau$ this joint probability will have a *LD* form, generalized (8):
$$P_\tau(\vec{M}) \sim e^{-\tau \Phi(\vec{M}/\tau)}, \tag{22}$$

where the *LD* function now depends on the whole vector of intensive observables $(M_1/\tau;...;M_N/\tau)$. The corresponding moment generating function $\vec{M}$ also has a *LD* form at large $\tau$ [28]:

$$Z_\tau(\vec{s}) \equiv \sum_{\vec{M}} e^{-\vec{s}\cdot\vec{M}} P_\tau(\vec{M}) \sim e^{\tau\Theta(\vec{s})}, \qquad (23)$$

where for each observable $M_n$ there is a counting field $s_n$, collected in the vector $\vec{s} \equiv (s_1,...,s_N)$, and where the *LD* function $\Theta(\vec{s})$ is a now function of this whole vector. The *LD* function $\Theta(\vec{s})$ is the largest eigenvalue of the deformed master operator (2) $W_{\vec{s}}$ (see below before (26)) [32]. Eq. (23) and the equation for the deformed master operator define an $(\tau;\vec{s})$-ensemble for a general set of dynamical order parameters $\vec{M}$ [32]. There is an equivalent construct for studying the statistics of $\vec{M}$ in trajectories where the total activity $K$ is fixed. The probability of observing $\vec{M}$, together with a total time $\tau$, for a fixed and large $K$, takes the form, generalized (17):

$$P_K(\tau,\vec{M}) \sim e^{-K\Phi(\tau/K,\vec{M}/K)}. \qquad (24)$$

The corresponding moment generating function is:

$$Z_K(x,\vec{s}) \equiv \sum_{\vec{M}} \int_0^\infty d\tau e^{-x\tau - \vec{s}\cdot\vec{M}} P_K(\tau,\vec{M}) \qquad (25)$$

$$\sim e^{KG(x,\vec{s})}.$$

Average values are determined the same way as in expressions (12), (18), $\frac{\langle\tau\rangle}{K} = -\frac{\partial}{\partial x}G(x,\vec{s})|_{\vec{s},x=0}$, $\frac{\langle M_i\rangle}{K} = -\frac{\partial}{\partial s_i}G(x,\vec{s})|_{x,\vec{s}=0}$. Equations (24) and (25) define a generalized $x$-ensemble [32]. In the case where $\vec{M}$ corresponds only to the activity $K$ the function $G(x,s) = g(x) - s$, and Eqs. $\Theta(\vec{s}) = x_*(\vec{s})$, $G(x_*(\vec{s}),\vec{s}) = 0$ reduce to (21).

As in [9], suppose that the system starts in an initial state $\psi$. For a random waiting time $t_w$ the system evolves continuously so that its unnormalized state at time $t < t_w$ is $exp(-iH_{eff}t)\psi$, where $H_{eff}$ is a non-selfadjoint effective Hamiltonian. At the time of detection, a click with label $i = 1,...,N_L$ is recorded, and the system's conditional state is updated by applying a jump operator $L_i$. A full detection process is given by a finite measurement trajectory $\mathbf{X}=((t_1, i_1),...,(t_n, i_n))$, where $0 \leq t_1 \leq \cdots \leq t_n$. Each such trajectory has the final time $T[X]=t_n$, and a total number of jumps $K[X]=n$. There are some natural ways of obtaining finite trajectories. The first one repeats the single detection process $K$ times. This scheme has an associated state transformation given by $J[\mathbf{X}] = J(t_{w,n},i_n)...J(t_{w,2},i_2)J(t_{w,1},i_1)$, where $t_{w,n} = t_n - t_{n-1}$, $t_{w,1} = t_1$ are the waiting times between the detection events or quantum jumps. In other words, given a trajectory $X$ resulting from this process, the system is at the end in state $J[\mathbf{X}]\psi$; $J[t_w,i] = L_i e^{-it_w H_{eff}}$ is the jump operator affecting the total change $\psi \to J[t_w,i]\psi$.

We are interested in quantities obtained by incrementing with some amount at the addition of each particle. Such a quantity is of the form $F[X]=\sum_{n=1}^{K[X]} F(t_{w,n},i_n)$, where $F(t_w, i)$ is a (possibly vector-valued) quantity depending only on a single waiting time $t_w$, and some property of the system we are monitoring, say "spin" $i$. The most important ways of truncating an infinite trajectory, which lead to statistical ensembles of field particles, are with either fixed dynamic activity $K$ and fluctuating $\tau$, or fixed the time $\tau$ and fluctuating $K$, obtained by taking $F(t_w, i) = t_w$ and $F(t_w, i) = 1$, respectively. We can also measure a "spin" operator $\hat{M}$ corresponding to $F(t_w, i) = M(i)$, where $M$ is some (in general vector-valued) quantity depending on $i$. The associated probability distributions are given by $P_K(T, M) := tr[\rho^{FC}{}_K \delta((T, M) - (\hat{T}, \hat{M}))]$, where $\rho^{FC}{}_K$ is the reduced density matrix on the output alone of the output space $F^{out}$ [9]. The associated generating functions are given by $Z_K(x, c) := tr[\rho^{FC}{}_K e^{-x\hat{T} - c\hat{M}}] = tr[\rho T^K{}_{x,c}(I)]$, written in terms of the deformed generators $T_{x,c}$ and $W_{s,c}$ [9] obtained from $T$ and $W$ (2) by replacing $J[X]$ and $V_\tau[X]$ with $J^\tau_{x,c}[X]$ and $V^\tau_{s,c}[X]$ [9], where [32] $T_{x,\vec{s}} = \sum_{C`\neq C} \frac{w_{\vec{s}}(C \to C`)}{x + R(C)}|C`\rangle\langle C|$, $w_{\vec{s}}(C \to C`)$ are the coefficient of the off-diagonal entries of $W_{\vec{s}} = \sum_{C'\neq C} e^{-\vec{s}\cdot\vec{m}(C \to C')} w(C \to C')|C'\rangle\langle C| - \sum_C R(C)|C\rangle\langle C|$, $m_n(C \to C')$ is the change in $M_n$ under the transition $C \to C'$ (for the activity this was just 1 for all $C, C'$ as it counted all transitions equally):

$$T_{x,c} = (xId + R)^{-1}(\sum_{i=1}^{N_L} e^{-c\cdot M(i)} L^\dagger_i(\cdot) L_i), \qquad (26)$$

the inverse $(xId + R)^{-1} = \int_0^\infty dt (e^{-it(H_{eff} - ix/2)})^\dagger (\cdot) e^{-it(H_{eff} - ix/2)}$ exists whenever $\|e^{-it(H_{eff} - ix/2)}\| \leq 1$, which holds for all $x > x_{min}$, where $x_{min} = 2Im\lambda_0 \leq 0$ and $\lambda_0$ is the eigenvalue of $H_{eff}$ with the maximum imaginary part. The elements of $T_x$ represent the Laplace transforms of the factors in the integrand of (16), since in (18) and (25) $Z_K(x)$ is the Laplace transform of $P_K(\tau)$. Then the *LD* function $g(x)$ corresponds to the logarithm of the largest eigenvalue $T_x$. We restrict to $x > x_{min}$ subsequently; $H_{eff}$ is a non-selfadjoint *effective Hamiltonian* (for open quantum systems), $H_{eff} = H - (i/2)\sum_m L^\dagger_m L_m$, where $H$ is a selfadjoint operator interpreted as the system's Hamiltonian when isolated from the environment.

Similar expressions are written in [9] for $W_{s,c}$, $P_\tau(K, M)$, $Z_\tau(s, c)$ for fixed volume $T$ and fluctuating $K$. We now suppose that $T_{x,c}$ has a unique eigenvalue $e^{g(x,c)}$ equal to its spectral radius, and that $W_{s,c}$ has a unique eigenvalue $\theta(s,c)$ with largest real part.

We assume, that $\vec{M} = K, M_1$, where $M_1$ is some quantity depending on the "spin" operator $i$ [9]. We write the distribution of the values $\tau, K, M_1$ in the form:

$$P_{s,x,c}(\tau, K, M_1) = e^{-sK} e^{-x\tau - cM_1} P(\tau, K, M_1) / Z(s, x, c), \qquad (27)$$

where

$$Z(s, x, c) = \sum_{M_1} \sum_K \int_0^\infty d\tau e^{-sK} e^{-x\tau - cM_1} P(\tau, K, M_1). \qquad (28)$$

Distributions of the form (27)-(28) were obtained, for example, in [87]. In [52, 53, 83-84] the joint distribution of *FPT* and system energy was obtained. Similar operations were carried out in [88], where distributions were obtained whose form is closer to the results of [52, 83-84] than to (29).

The value of $M_1$ depends on the "spin" operator [9] $i_n$ in $(t_{w,n}, i_n)$. Let us assume that the value of $M_1$ is fixed at the value $i_n$. In this case, the value of $K$ is also fixed. We assume that the values of $i_n$, $M_1$, and $K$ are sufficiently large for the relation $Z_K(x,c) \sim e^{g(x,c)}$ to hold. At fixed values of $M_1$, the factor $e^{-cM_1}$ in the denominator and numerator (27) is reduced, the relation is fulfilled:

$$\sum_{M_1} P(\tau, M_1, K) = P(\tau, K)$$
$$= P(\tau|K)P(K) = P_K(\tau)P(K),$$

where the conditional probability formula equal $P(\tau, K) = P(\tau|K)P(K)$, $P(\tau|K) = P_K(\tau)$ to is used. We equate the conditional probability $P(\tau|K)$ with the probability $P_K(\tau)$ from (16)-(18). Thus, the value of $K$ is fixed in the conditional probability, but then these fixed values of $K$ are averaged. It is assumed that there is an ensemble of systems with fixed values of $K$. Integration over $\tau$ in (28) leads at large $K$ to the expression (18), (19), $\int_0^\infty d\tau e^{-x\tau} P_K(\tau) = e^{Kg(x)}$, deformed generators [9] $T_{x,c}$ (26) act, and partition function (28) takes the form (7)-(9):

$$Z = Z_\tau(s_{ef}) =$$
$$\sum_K e^{-K(s-g(x))} P(K), \quad s_{ef} = s - g(x). \tag{29}$$

From equality (29) we obtain that $s_{ef} = s - g(x) = 0$, by (21). But for $s_{ef}=0$, the values of $s$, $g(x)$, and $\gamma$ are not equal to zero. We consider expression (29) as a formal relation; for small values of $\tau$ (values of $K$ are large, the *LD* function $g(x)$ is already used), when the equations for the eigenvalues of a transfer matrix operator $W_{s,c}$ [9, 32] should take into account other eigenvalues, except for the largest ones, and $s_{ef} = s - g(x) \neq 0$. From expression (29) we find the average values and second moments, and then, after differentiation, we set $s_{ef} = s - g(x) = 0$.

This corresponds to the fact that we pass to the limit of large values of the time $\tau$ after taking derivatives and determining the moments, as in the non-equilibrium statistical operator (*NSO*) method [89–91]. In the *NSO* method, calculations are also made for large volumes, and then the passage to the limit $\varepsilon \to 0$ is carried out. In [92], it was shown that $\varepsilon = 1/\bar{T}$, where $\bar{T}$ is the average lifetime of the system, analog of quantity $\tau$. Also, in expressions (30)-(34), at first large, but finite values of $K$, corresponding to a large volume, are chosen, and at the end - large values of $\tau$, corresponding to large *FPT* $\bar{T}$. Hence, using relation (29), we obtain expressions for the average values and second moments:

$$\langle K \rangle = -\frac{\partial \ln Z_{s_{ef}}}{\partial s_{ef}} \frac{\partial s_{ef}}{\partial s}\bigg|_{s_{ef}=0} = \langle K_0 \rangle. \tag{30}$$

From (18)-(19) we have $\ln Z_K(\gamma) = Kg(\gamma)$,

$$\langle \tau_\gamma \rangle = -K\frac{\partial g(\gamma)}{\partial \gamma}, \quad K = \langle K_0 \rangle, \tag{31}$$

$$D_K = \langle K^2 \rangle - \langle K \rangle^2 = \frac{\partial^2 \ln Z_{s_{ef}}}{\partial s^2}\bigg|_{s_{ef}=0} =$$
$$D_{K_0} = \langle K_0^2 \rangle - \langle K_0 \rangle^2, \tag{32}$$

$$D_\tau = \langle\tau^2\rangle - \langle\tau\rangle^2 = \frac{\partial^2 \ln Z_\gamma}{\partial \gamma^2} = K \frac{\partial^2 g(\gamma)}{\partial \gamma^2}. \tag{33}$$

The derivative of $lnZ(s_{ef})$ (29) concerning $\gamma$ leads to the identity:

$$\langle\tau\rangle = -\frac{\partial \ln Z(s_{ef})}{\partial \gamma} = -\frac{\partial \ln Z(s_{ef})}{\partial s_{ef}} \frac{\partial s_{ef}}{\partial \gamma} =$$

$$\langle K_0\rangle \frac{\langle\tau\rangle}{K}, \quad \frac{\partial s_{ef}}{\partial \gamma} = -\frac{\partial g(\gamma)}{\partial \gamma}, \quad \langle K_0\rangle = K.$$

Mixed derivatives of $lnZ(s_{ef})$ (29) allow one to obtain a correlation between the parameters $K$ and $\tau$:

$$D_{K\tau} = \langle K\tau\rangle - \langle K\rangle\langle\tau\rangle = \frac{\partial^2 \ln Z_{s_{ef}}}{\partial s \partial \gamma} =$$

$$\frac{\partial}{\partial \gamma}\bigg|_s \left(\frac{\partial \ln Z_{s_{ef}}}{\partial s}\bigg|_\gamma\right) = \frac{\partial^2 \ln Z_{s_{ef}}}{\partial s^2} \frac{\partial s_{ef}}{\partial \gamma} = D_{K_0} \frac{\langle\tau_\gamma\rangle}{K}. \tag{34}$$

This approach, together with relations (21), makes it possible to find the parameters $\tau$ and $K$ that are undefined in expressions (10) and (19): $\tau = \tau_0 = \langle\tau_{\gamma=0}\rangle$, $K = K_0 = \langle K_{\gamma=0}\rangle$. As shown in [93], from the theory of random processes it follows that the value $K = K_0$ is equal to the value of the boundary that the *FPT* of a random process with a fixed value of $K$ reaches. You can write expressions for higher-order correlators.

Let us check the fulfillment of the inequality $\langle\tau\rangle_A / \sqrt{\text{var}(\tau)} \leq \sqrt{K_A}$ obtained in [7] for expressions (30)-(34). In the notation of this article $\langle\tau\rangle_A = \tau_0$, $\text{var}(\tau) = D_\tau\big|_{\gamma=0}$, $K_A = K_0$. For the two-level models discussed below and for the three-level model, strict inequality is satisfied.

## 4. Dependences of the First and Second Moments of Random Variables of Activity and FTP on Changes in Entropy

All real physical processes modeled by stochastic processes occur with a change in entropy. These changes play a crucial role in calculating the average first-passage time (*FPT*) [83-86]. The average *FPT*, calculated at the zero value of the Laplace transform's argument of the *FPT* distribution, typically used to determine the average *FPT*, does not account for the influence of real processes.

Expressions (30)-(34) include the parameter $\gamma$, which equals $x$ in [2, 32], as part of the Laplace transform argument in (18), (25), (28), and (39). In this section, $\gamma$ is associated with the total entropy change in the system during the *FPT*. The moments of the random variables *FPT* and $K$ (mean values, variances, correlations) are expressed through $\gamma$ and the total entropy change during the *FPT*. We will now describe the algorithm for expressing moments (30)-(34) through the total entropy change.

The total entropy change $\Delta s_{tot}$ consists of the system's entropy change $\Delta s_{sys}$ and the entropy exchange with the environment $\Delta s_m$, where $\Delta s_{tot} = \Delta s_{sys} + \Delta s_m$. Expressions (30)-(34) depend on $\gamma$, and the values of $\Delta s_{sys}$ and $\Delta s_m$ also depend on $\gamma$. We can write the relation $\Delta s_{tot} = \Delta s_{sys}(\gamma) + \Delta s_m(\gamma)$, treating it as an equation for $\gamma$, which depends on the total entropy change $\Delta s_{tot}$. By solving this equation, we obtain $\gamma(\Delta s_{tot})$ and substitute it into expressions (30)-(34).

In this article, the same thermodynamic quantity, which is conjugate to a random *FPT*, is represented with different notations, $\gamma$ in expressions (21), (31)-(34), and $x$ in expressions (18)-(20), (27)-(29). This is due to an attempt to connect the results for *x*-ensembles in [2, 32], where the notation $x$ is used, with the results in [52-53], [83-84], where $\gamma$ is used. In both cases, *FPT* is treated as a random thermodynamic parameter, and $\gamma$ (or $x$) is conjugate to this thermodynamic variable, as in the distributions (20), (27), and (39).

In expression (19), the function $g(x)$, an *LD* function depending on $x$ from [2, 32], can be written as $x=\gamma$ using the notation from [83-84]. According to relation (21), the argument $\gamma$ is related to the cumulant $\theta(s)$. We express this argument $\gamma$ in terms of the entropy change over the *FPT*. The value of $K$ from (19) is proportional to the average $K$ from (12). Setting $s=0$, we assume that the random process $\tau$ changes accordingly, denoted as $\tau_0$ at $s=0$. For $s \neq 0$, the dependence of the parameter $K$ on $\gamma$ takes the form:

$$\langle K \rangle = \langle K_s \rangle = \langle K_\gamma \rangle = -\frac{\partial \theta(s)}{\partial s}\bigg|_{s=g(\gamma)} \tau. \tag{35}$$

Subsequent calculations will be performed for a two-level model. In [32-33], a classical two-level system is described. The operator $W_s$ is given by $W_s = \begin{pmatrix} -\eta & e^{-s}\kappa \\ e^{-s}\eta & \kappa \end{pmatrix}$, where $W_s = \sum_{C' \neq C} e^{-s} w(C \to C')|C'\rangle\langle C| - \sum_C R(C)|C\rangle\langle C|$ [2, 8], $W_s$ is a deformed operator $W(2)$, and $\theta(s)$ (36) is its largest eigenvalue; $\eta$ and $\kappa$ are transition rates. In a two-level system with only two configurations, $C \in \{0,1\}$, and the transition rates are $w(0 \to 1) = \kappa$ and $w(1 \to 0) = \eta$. From $\theta(s)$, the activity can be calculated. The average activity per unit time can be expressed in terms of (35)-(37) as expected. For the case $\kappa = \eta$, the *LD* function simplifies to $\theta(s)=\eta(e^{-s/2}-1)$, which is the cumulant generating function for a Poisson process with rate $\eta$.

The operator $T_x$ (26) for this problem is given by $T_x = \begin{pmatrix} 0 & \kappa/(x+\kappa) \\ \eta/(x+\eta) & 0 \end{pmatrix}$, and from its largest eigenvalue, we obtain the *LD* function $g(\gamma)$ (36). This function $g(\gamma)$ is the inverse of the function $\theta(s)$ (36). The moments of the total time $<\tau>$ are obtained from $g(\gamma)$. In particular, the average total time, scaled by the number of jumps, is $\langle \tau \rangle_K / K = -g'(0) = (\kappa+\eta)/\kappa\eta$, which is the inverse of (37). Analogous relations between the moments of $K$ in the fixed $\tau$ ensemble and those of $\tau$ in the fixed $K$ ensemble can be obtained from Eq. (21).

For a classical two-level system [32, 33] as described in [33], the following expressions are obtained:

$$g(\gamma) = -\ln(1 + \gamma a_1 + \gamma^2 a_2), \quad a_1 = \frac{\eta+\kappa}{\eta\kappa}, \quad a_2 = \frac{1}{\eta\kappa};$$

$$\theta(s) = \frac{1}{2}[\sqrt{(\eta-\kappa)^2 + 4\eta\kappa e^{-s}} - (\eta+\kappa)]. \tag{36}$$

From (10), (12) we obtain:

$$\frac{\langle K_0 \rangle}{\tau} = \frac{1}{a_1} = \frac{\eta\kappa}{\eta+\kappa} = -\frac{\partial \theta(s)}{\partial s}\bigg|_{\gamma=0, s=0}. \tag{37}$$

We will limit our calculations to the case $K = K_0 = \langle K_{\gamma=0} \rangle$, though it is possible for $K$ to depend on $\gamma$ (see Appendix A).

## 4A. Example: Classical Two-Level System, Partition Function (39)

Let's consider the entropy change. The total entropy change $\Delta s_{tot}$ consists of the system's entropy change $\Delta s_{sys}$ and the exchange of entropy with the environment $\Delta s_m$:

$$\Delta s_{tot} = \Delta s_{sys} + \Delta s_m. \tag{38}$$

The Gibbs/Shannon entropy for a distribution of the form (27)-(28) [52-53], [83-84], [94], [88]:

$$p_{s\gamma} = e^{-sK-\gamma\tau_\gamma} / Z_{s\gamma}, \quad Z_{s\gamma} = Z(s,\gamma) = Z_s Z_\gamma, \tag{39}$$

$$Z_s = Z_\tau(s) = \sum_K e^{-sK} P_\tau(K) \sim e^{\tau\theta(s)},$$

$$Z_\gamma = Z_K(\gamma) = \int_0^\infty d\tau e^{-\gamma\tau} P_K(\tau) \sim e^{Kg(x)}$$

is given by:

$$s_\gamma = s_{sys} = -\langle \ln p_{s\gamma} \rangle = s\langle K_\gamma \rangle + \gamma\langle \tau_\gamma \rangle + \ln Z_{s\gamma}, \tag{40}$$

where the average values are:

$$\langle K_\gamma \rangle = -\partial \ln Z_s / \partial s = \langle K_{s=g(\gamma)} \rangle. \tag{41}$$

Entropy (40) includes the total average values of $<K_\gamma>$, and $<\tau_\gamma>$ not the ratios $<K_\gamma>/\tau$ and $<\tau_\gamma>/K$. The quantities $lnZ_s$ and $lnZ_\gamma$ also depend on the parameters $\tau$ and $K$, complicating the problem due to the uncertainty in the expressions for $K$ and $\tau$ (as discussed in the paragraph following equation (34)).

The Gibbs/Shannon entropy (40) is naturally generalizable to non-equilibrium states, as it remains well-defined even when $p(x)$ is not the Boltzmann distribution [94], [95]. The Clausius relation $\Delta S \geq -Q/T$ (where $Q$ is heat) relates the change $\Delta S$ in the system's entropy to the heat $Q$ exhausted into an ideal thermal reservoir at temperature $T$. This relation should hold for this choice of entropy in a broad class of Markovian stochastic processes on a finite set of states [96].

Expressions (42)-(45) were obtained in [10] from the Gibbs (Shannon) entropy. There are different types of entropy, for example, thermodynamic entropy and information entropy. While these types are not identical, many works, such as [89-91], [94-96], [101-103], draw analogies between them and effectively use their similarities. In general, the formulation of the entropy production problem is not universal. It depends on the dynamic laws governing the system and the underlying physical system itself. We will assume that for the entropy inside the system, expressions (40) and (A.3) are valid, and for the exchange of entropy with the environment, expressions (42)-(45) apply.

The partition function (39) differs from expression (29) and is derived from expression (28) if the value of $K$ is taken from the conditional probability $P(\tau|K)$ and $e^{Kg(x)}$ is not averaged over $K$ (assuming, for example, that $K$ is equal to the mean value), as was done in [32]. However, the approach leading to expression (29) is more consistent. Quantity (40) is equal to the system's entropy $s_{sys}$, the value $\langle \tau_\gamma \rangle$ from (31) and (A.2). A distribution of the form (39) was obtained in [88] as an *sx*-ensemble. This distribution differs from those obtained in [52, 83-84] by replacing the pair of conjugate quantities $\beta u$ (the product of inverse temperature $\beta$ and energy $u$) with $sK$ (activity $K$ and the conjugate field $s$).

In [97], changes in $K$-activity are associated with entropy production within the system and with the exchange of entropy between the system and its environment. For a single transition $C \to C'$, the change in entropy is:

$$\Delta s_1(C,C') = \ln[w(C \to C')/w(C' \to C)]. \tag{42}$$

where $w(C \to C')$ is the rate of the jump process, the change in entropy during the exchange with the medium while moving along the trajectory $Y_K(C_0 \to C_1 \to ... \to C_K)$ is given by:

$$\Delta s_m[C(t)] = \sum_{\alpha=1}^{K} \Delta s_1(C_{\alpha-1}, C_\alpha), \tag{43}$$

as $K$ changes from $K_0$ to $K_\gamma$:

$$\Delta s_{m(K_0 K_\gamma)}[C(t)] = \sum_{\alpha=K_0}^{K_\gamma} \Delta s_1(C_{\alpha-1}, C_\alpha), \tag{44}$$

where we sum over all configuration changes. The corresponding dynamical partition function is [10, 97]:

$$Z(\lambda,\tau) = \left\langle e^{-\lambda s_m} \right\rangle \sim e^{\tau \theta(\lambda)}, \tag{45}$$

where $\lambda$ is the parameter conjugate to $s_m$. Analogous to the activity, the mean entropy production rate in the $\lambda$-ensemble is given by $\langle s_m \rangle / \tau = -\partial \theta(\lambda)/\partial \lambda$. Assuming that the quantity $\Delta s_1 = \Delta s_1(C_{\alpha-1}, C_\alpha)$ is constant, expression (45) takes the form:

$$\Delta s_{m(K_0 K_\gamma)}[C(t)] = \Delta s_1(\langle K_\gamma \rangle - \langle K_{\gamma=0} \rangle). \tag{46}$$

Using expression (39) and the results of [98] (43), we obtain for the term from (38):

$$\langle s_m \rangle = -\tau \partial \theta(\lambda)/\partial \lambda, \quad \tau = \tau_0,$$
$$\Delta \langle s_m \rangle = \langle s_{m|K_\gamma} \rangle - \langle s_{m|K_0} \rangle = \tag{47}$$
$$-\tau[\partial \theta(\lambda)/\partial \lambda \big|_{\lambda=\gamma} - \partial \theta(\lambda)/\partial \lambda \big|_{\lambda=0}].$$

Expression (47) coincides with (46) at $\Delta s_1 = (+/-)1$ [60]. Expressions (42)-(45) were derived in [97] for the thermodynamics of trajectories, expanding the $s$-ensemble approach to driven systems, based on the results from [10]. In [98], the difficulties in the physical interpretation of general definitions of entropy production for specific physical systems were noted. In both [98] and [99], entropy production was obtained for the Markov jump process. Expressions (42)-(45) for dynamics equations (1)-(2) were used in [100] to connect gauge invariance in stochastic dynamics with fluctuation theorems. These issues are also discussed in [101-103] and other works.

In [94], the "microscopic reversibility relation" plays an important role in terms of the thermodynamic entropy of the environment, as it encodes the time-reversibility of full microscopic dynamics in coarse-grained stochastic dynamics. In general, the formulation of the entropy production problem is not universal, as it depends on the dynamic laws governing the system as well as the underlying physical system.

Let us express the parameter $\gamma$ in terms of the change in entropy. Expanding the expressions from relation (36) into a series in $\gamma$ up to the power $\gamma^2$:

$$\ln(1 + \gamma a_1 + \gamma^2 a_2) = \gamma a_1 - \gamma^2 c_3 + ...,$$
$$c_3 = a_1^2/2 - a_2, \tag{48}$$
$$\frac{a_1 + 2\gamma a_2}{1 + \gamma a_1 + \gamma^2 a_2} = a_1 - \gamma 2 c_3 + \gamma^2 a_1 (2c_3 - a_2) + ....$$

If in (40) we used expansions (48), then:
$$\Delta\langle s_{sys}\rangle/\langle K_0\rangle = (\langle s_0\rangle - \langle s_\gamma\rangle)/\langle K_0\rangle =$$
$$\gamma^2[2a_2 - a_1^2] = \gamma^2 a_s, \tag{49}$$

where $a_s = [2a_2 - a^2{}_1]$

For further estimates, we need to know the explicit form of the function $g(\gamma)$. Consider a classical two-level system [32, 33] (a quantum system can also be considered), where $g(\gamma)$ takes the form (36). From (47) and (48), up to $\gamma^2$, we obtain:
$$\langle \Delta s_m \rangle = \gamma^2(\pm a_m) + \gamma(\pm b_m), \tag{50}$$

where $a_m = 4a^2{}_2/a^2{}_1 - a_2$, $b_m = a_1 - 2a_2/a_1$, the sign depends on whether the flows enter the environment (+) or exit the system (-). The plus or minus sign is determined by whether the flow is directed into the system from the environment or from the system to the environment.

Using expression (36) leads to a transcendental equation for the parameter $\gamma$. If we expand into a series and restrict ourselves to quadratic terms in $\gamma$, then from (38), (49), and (50) we obtain an equation for $\gamma$ of the form:
$$a\gamma^2 + b\gamma + d = 0, \quad a = -(a_s \pm a_m),$$
$$b = -(\pm)b_m, \quad d = \Delta\langle s_{tot}\rangle/\langle K_0\rangle. \tag{51}$$

The coefficients of this equation depend on the sign of expressions (50), and (51). The solution of equation (51) takes the form:
$$\gamma = [\pm\sqrt{b^2 - 4da} - b]/2a, \tag{52}$$

where $a$ and $b$ are parameters from expressions (50) and (51) that depend on the sign (+) or (-):
$$a_+ = -a_s - a_m, \quad a_- = -a_s + a_m, \tag{53}$$

where $a_+ = -(a_s + a_m) = a_1^2 - a_2 - 4a^2{}_2/a^2{}_1$, $a_- = -(a_s - a_m) = a_1^2 - 3a_2 + 4a^2{}_2/a^2{}_1$, $b_+ = -b_m$, $b_- = b_m$, $b_+ < 0$, $b_- > 0$, as for model (36) $2a^2{}_1 - 5a_2 > 0$, $7a_1/2 + 3a_2/a_1 > 0$.

If in (52) we choose the + sign in front of the square root, then we get two solutions depending on the signs in (51). Let's denote them:
$$\gamma_{1(+)(+)} = [\sqrt{b_+^2 - 4da_+} - b_+]/2a_+$$
$$= [\sqrt{1 - 4da_+/b_m^2} + 1]b_m/2a_+ > 0. \tag{54}$$

The first subscript (+) in (54) denotes the sign before the square root in (52), and the second plus sign corresponds to the sign chosen in (50)-(51). At $\Delta\langle s_{tot}\rangle = 0$, expression (54) equals $b_m/a_+ \neq 0$. Expressions (49) and (50) are not zero, but they yield $\Delta\langle s_m\rangle = -\Delta\langle s_{sys}\rangle$. This describes a stationary nonequilibrium state (at $\Delta\langle s_{tot}\rangle = 0$) as defined in [104]. In [105], it is shown that $\gamma \sim \sigma_s$, entropy production in the system $\sigma_s \sim \Delta\langle s_{sys}\rangle/\langle \tau_\gamma\rangle$ is positive. The system has flows $q$, $\gamma \sim q$ with the possibility of nonequilibrium stationary states when $\Delta\langle s_{tot}\rangle = 0$, but $\Delta\langle s_{sys}\rangle \neq 0$, $\Delta\langle s_m\rangle \neq 0$. For the "-" sign in (51), we obtain the solution:

$$\gamma_{1(+)(-)} = [\sqrt{1-4da_-/b_m^2} - 1]b_m/2a_-,$$
$$\Delta\langle s_{tot}\rangle \geq 0. \tag{55}$$

At $\Delta\langle s_{tot}\rangle = 0$, expression (55) vanishes. Expressions (49) and (50) also vanish. This corresponds to the equilibrium state (at $\Delta\langle s_{tot}\rangle = 0$).

If we choose the "-" sign in front of the square root in (52), two solutions depending on the signs in (51)-(52) also arise:

$$\gamma_{2(-)(+)} = [-\sqrt{1-4da_+/b_m^2} + 1]b_m/2a_+, \tag{56}$$

$$\gamma_{2(-)(-)} = -[\sqrt{1-4da_-/b_m^2} + 1]b_m/2a_-, \quad \Delta\langle s_{tot}\rangle \geq 0. \tag{57}$$

For negative values of $\gamma$, the convergence condition of the Laplace transforms of the *FPT* distribution and the partition function $Z_\gamma$ from (39) must be satisfied. At large times, time distributions tend to an exponential distribution of the form $\tau_0^{-1}\exp(-\tau/\tau_0)$, $\tau_0 = \langle \tau_{\gamma=0}\rangle$.

For the Laplace transform of $\gamma$, an exponential distribution of the *FPT* with mean value $\tau_0$, $\gamma < 0$, must satisfy the condition:

$$\gamma + 1/\tau_0 > 0 \tag{58}$$

In case (57), for the convergence of $Z_\gamma$, the condition $\Delta\langle s_{tot}\rangle \geq 0$ must be satisfied.

For the parameter values chosen below, the values $\gamma_{(-)(-)}$ in equation (57) are not realized. In the case of equation (56), condition (58) imposes a constraint of the form $d<0.00676$ for the parameter values $\eta = 5$, $\kappa = 1.25$ as given in [33].

The condition for the positivity of the radical expression must also be satisfied, which imposes another boundedness condition on $d$ of the form $d < b^2/4a$. For the case of the positive sign in equation (50) and the given parameter values $\eta = 5$, $\kappa = 1.25$, this condition restricts $d$ to $d<0.1567$. For the case of the negative sign in equation (50), the restriction is $d<0.1863$.

If we substitute $\Delta\langle s_{tot}\rangle = 0$ into equation (51), then we obtain the equation for $\gamma$:

$$\gamma[\gamma(a_s \pm a_m) \pm b_m] = 0. \tag{59}$$

The first root of this equation is $\gamma=0$. Substituting this root into expressions (50) and (49) gives a result where the values remain unchanged, $\Delta\langle s_{sys}\rangle = \Delta\langle s_m\rangle = 0$. This situation describes the equilibrium state as per equations (55) and (56).

The second root of equation (59) is non-zero, $\gamma = \mp b_m/(\pm a_m + a_s)$. In this case, $\Delta\langle s_{sys}\rangle \neq 0$, $\Delta\langle s_m\rangle \neq 0$, $\Delta\langle s_{tot}\rangle = 0$; we obtain different values for $\gamma_{(+)(+)}$, leading to a stationary non-equilibrium state (54). State (57) is not realized.

From equation (36), we can observe that this case $\gamma = 0$, $s = 0$ corresponds to a phase transition, as described in [2, 12]. The system dynamics exhibit two phases: an active phase for $s<0$ and an inactive phase for $s>0$, also noted in [2, 12]. The model used in equation (36) suggests the existence of two equilibrium states, one of which (55) is possible within the range $d<0.1567$, while the second is possible only within the narrow range defined by condition (58).

Additionally, a stationary non-equilibrium state was obtained where certain values $\gamma = 0$, $s = 0$ are not reached, and no phase transition occurs. Therefore, the system exhibits an active phase (55) with specific parameter values $\gamma > 0$, $s < 0$ and an inactive phase $\gamma < 0$, $s > 0$ (56) that is more constrained. The transition between these phases occurs as $\gamma$ shifts from state (55) to (56) and back, passing through an equilibrium state where the values $s = 0$, $\Delta \langle s_{tot} \rangle = \Delta \langle s_{sys} \rangle = \Delta \langle s_m \rangle = 0$ remain unchanged. Moreover, there exists another active phase (54) $\gamma > 0$, $s < 0$ — a stationary non-equilibrium state without a phase transition.

If we consider not the average, but the random total change in entropy, this value can be negative, not exceeding the Boltzmann constant. In this case, the exchange of entropy with the surrounding medium should be treated as a random variable.

As in expressions (30)-(34), we now consider the observable dynamic activity K and the First Passage Time (*FPT*), as well as the first and second moments of these observables and their correlation. The *s*-tilted probability density of $x$ satisfies the relation $f_s(x) \sim e^{sx} f(x)$. The expressions for the moments depend on the parameter $\gamma$, which is related to the change in entropy. The equations of motion are also modified to produce a tilted or twisted equation of motion. The function $\theta(s)$ can be obtained by deforming the master operator $W$ (2) and replacing it with $W_s$ [10, 28]. In particular, for the case of activity, this deformed operator is discussed in [2, 8, 12], and for *FPT*, the deformed generators are given by $T_{x,c}$ (26) (with $x=\gamma$). The scaled cumulant generating function $\theta(s)$ and the rate function $\varphi(a)$ (8) are related by a Legendre transform $\varphi(a) = -min_s [\theta(s) + sa]$ (13), such that $\varphi(a) = -\theta[s(a)] - s(a)a$, with parameters $s$ and $a$ related by $a(s) = -\theta'(s)$.

Let us now illustrate the results obtained concerning entropy changes. We will outline the ratios according to which Figures 1-3 were constructed. A two-level model (36) is employed with parameter values $\eta = 5$, $\kappa = 1.25$ as given in [33]. To avoid ambiguity in determining the values of $K$ and $\tau$ in equations (39)-(41), we assume $K = K_0$. Let $K_0 = 100$. From equations (30)-(34), we then have the following for $K = \langle K_\gamma \rangle \approx \langle K_0 \rangle$, $a_1 = 1$, $a_2 = 0.16$, $\langle K_0 \rangle = 100$:

$$\langle T \rangle = \langle \tau_\gamma \rangle \simeq \langle K_0 \rangle \frac{\partial g(\gamma)}{\partial \gamma}$$

$$= \langle K_0 \rangle \frac{a_1 + 2\gamma a_2}{1 + \gamma a_1 + \gamma^2 a_2} = 10^2 \frac{1 + 0.32\gamma}{1 + \gamma + 0.16\gamma^2}, \quad (60)$$

$$D_{K_0} = \frac{\eta \kappa (\eta^2 + \kappa^2)}{(\eta + \kappa)^3} \langle T_0 \rangle =$$

$$0.68 \langle T_0 \rangle = 0.68 \langle K_0 \rangle, \quad \langle T_0 \rangle = \langle K_0 \rangle a_1,$$

$$D_T = \langle K_0 \rangle \frac{2(-a_2 + a_1^2/2 + \gamma a_1 a_2 + \gamma^2 a_2^2)}{(1 + \gamma a_1 + \gamma^2 a_2)^2}, \quad (61)$$

$$CorrTK = \langle TK \rangle - \langle T \rangle \langle K \rangle$$

$$= 0.68 \langle K_0 \rangle \frac{a_1 + 2\gamma a_2}{1 + \gamma a_1 + \gamma^2 a_2}, \quad (62)$$

where $\langle T \rangle = \langle \tau \rangle$ is the average *FPT*, $D_{K_0} = \langle K_0^2 \rangle - \langle K_0 \rangle^2$ is the variance of $K_0$, $D_T = \langle T^2 \rangle - \langle T \rangle^2$ is the variance of *FPT*, and *CorrTK* is the correlation between *T* and *K*, as given by equation (34). The value of $\gamma$ in equations (60)-(62) can take the forms (54)-(56). For the model (36) with the chosen parameter values $\eta = 5$, $\kappa = 1.25$, the values in equations (54)-(56) are:

$$\gamma_{1(+)(+)} = 0.461[\sqrt{1 - 6.381d} + 1],$$
$$\gamma_{2(-)(+)} = 0.546[\sqrt{1 - 5.384d} - 1], \quad (63)$$
$$\gamma_{1(+)(-)} = 0.461[1 - \sqrt{1 - 6.381d}], \quad d = \Delta \langle s_{tot} \rangle / \bar{K}_0.$$

The condition for the radical expression to remain positive imposes a limit on the quantity *d*, such that $d < 1/6.381 = 0.1567$. Figures 1-3 depict the behavior of the average *FPT* $\langle T(d) \rangle$ (60), the variance of *FPT* (61), and the correlator *CorrTK* (62) as functions of $\gamma_{1(+)(+)}$ (Fig. 1), $\gamma_{1(+)(-)}$ (Fig. 2), $\gamma_{2(-)(+)}$ (Fig. 3) (63), and parameter $d = \Delta \langle s_{tot} \rangle / \langle K_0 \rangle$.

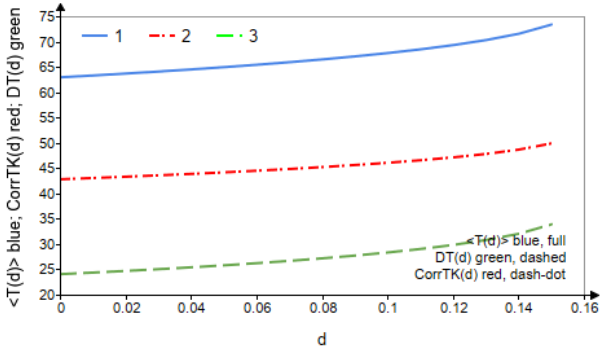

Fig. 1: Behavior of $<T(d)> = \langle T \rangle = \langle \tau_\gamma \rangle$ (sec) (60) [solid line, blue], $D_T(d)$ (sec²) (61) [dashed line, green], and *CorrTK(d)* (sec) (62) (62) [dot-dash, red] for $\gamma_{1(+)(+)}$ (54) and (63), plotted against $d = \Delta \langle s_{tot} \rangle / \langle K_0 \rangle$ in the range $0 < d < 0.156$.

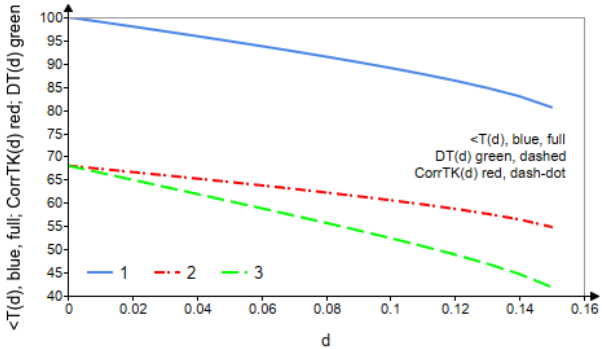

Fig. 2: Behavior of $<T(d)> = \langle T \rangle = \langle \tau_\gamma \rangle$ (sec) (60) [solid line, blue], $D_T(d)$ (sec²) (61) [dashed line, green], and *CorrTK(d)* (sec) (62) [dot-dash, red] for $\gamma = \gamma_{1(+)(-)}$ (55) and (63), plotted against $d = \Delta \langle s_{tot} \rangle / \langle K_0 \rangle$ in the range $0 < d < 0.156$.

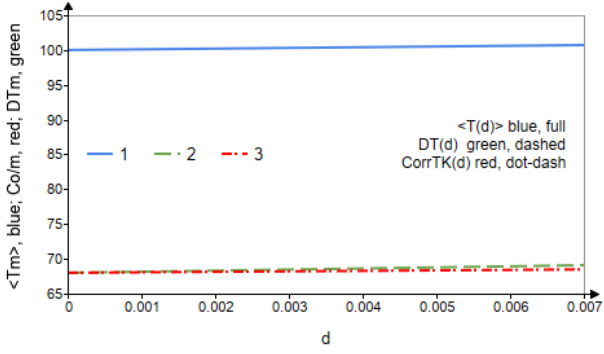

Fig. 3: Behavior of $<T(d)> = \langle T \rangle = \langle \tau_\gamma \rangle$ (60) [solid line, blue], $D_T(d)$ (61) [dashed line, green], and $CorrTK(d)$ (62) [dot-dash, red] for $\gamma = \gamma_{2(-)(+)}$ (56) and (63), plotted against $d = \Delta \langle s_{tot} \rangle / \langle K_0 \rangle$ in the range $0<d<0.00676$, corresponding to the convergence of the Laplace transform of the FPT distribution $Z_\gamma$, obtained from equation (58).

We express the dependence of the average activity $\langle K_\gamma \rangle$ on the parameter $\gamma$ in the form $\langle K_\gamma \rangle \approx -\langle \tau_0 \rangle \frac{\partial \theta(s)}{\partial s} \Big|_{s=g(\gamma)}$. Then:

$$\langle K_\gamma \rangle = \langle \tau_0 \rangle \frac{\eta \kappa (1 + a_1 \gamma + a_2 \gamma^2)}{(\eta + \kappa) \sqrt{1 + 4\eta\kappa(a_1 \gamma + a_2 \gamma^2)/(\eta+\kappa)^2}} = \langle K_0 \rangle \frac{1 + a_1 \gamma + a_2 \gamma^2}{\sqrt{1 + 4(a_1 \gamma + a_2 \gamma^2)/a_1(\eta+\kappa)}}. \quad (64)$$

Figure 4 shows the behavior of $\langle K_\gamma \rangle$ (64) as a function of $\gamma_{1(+)(+)}$ (63) $\langle K_{\gamma 1+} \rangle$, $\gamma > 0$, $s < 0$, in the active phase, $\gamma_{1(+)(-)}$ (63) $\langle K_{\gamma 1-} \rangle$, at $\gamma < 0$, $s > 0$ in the inactive phase, and $\gamma_{2(-)(+)}$ (63) $\langle K_{\gamma 2+} \rangle$ at $\gamma > 0$, $s < 0$ (active phase), plotted against $d = \Delta \langle s_{tot} \rangle / \langle K_0 \rangle$ in the range $0<d<0.156$. Similar dependencies (but with respect to $s$, rather than $\gamma$) were reported in [2, 12].

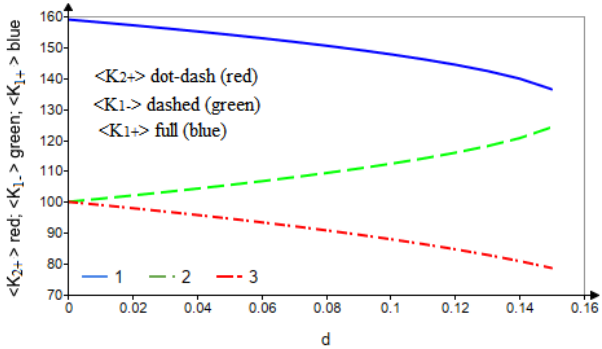

Fig. 4: Behavior of $\langle K_\gamma \rangle$ (64) depending on $d = \Delta \langle s_{tot} \rangle / \langle K_0 \rangle$ (63), shown as [solid line, blue], depending on $\gamma_{1(+)(-)}$ (63), $\gamma > 0$, $s < 0$ for the active phase [dashed line, green], and $\gamma_{2(-)(+)}$ (63), $\langle K_{\gamma 2+} \rangle$, $\gamma < 0$, $s > 0$, for the inactive phase [dot-dash, red]. The panel illustrates $\langle K_{\gamma 1-} \rangle$, $\langle K_{\gamma 2+} \rangle$ and $\langle K_{\gamma 1+} \rangle$ as functions of $d = \Delta \langle s_{tot} \rangle / \langle K_0 \rangle$ in the range $0<d<0.156$.

### 4B. Example: Classical Two-Level System, Partition Function (29)

The calculations for Figs. 1-4 were carried out using expression (39), with the partition function of the form $Z_{s\gamma} = Z(s,\gamma) = Z_s Z_\gamma$, $Z_s = Z_\tau(s) = \sum_K e^{-sK} P_\tau(K) \sim e^{\tau\theta(s)}$, as in equations (9), (10), $Z_\gamma = Z_K(\gamma) \equiv \int_0^\infty d\tau e^{-\gamma\tau} P_K(\tau)\big|_{x=\gamma} = e^{Kg(\gamma)}$, (18), (19). If we instead calculate using expression (29), where the partition function is equal to $Z = Z_\tau(s_{ef}) = \sum_K e^{-K(s-g(\gamma))} P(K)$, $s_{ef} = s - g(\gamma)$, and for large times $Z_\tau(s_{ef})_{\tau\to\infty} \to 1$, $\langle\tau\rangle_\gamma = -\langle K_0\rangle \partial g(\gamma)/\partial\gamma$, $\langle K\rangle_\gamma = \langle K\rangle_0$, as in equations (30), (31), then $\ln Z_\tau(s_{ef})_{s_{es}=0} = 0$. Using expression (47) to determine $\Delta s_m$ and (38), we obtain, using expressions (21) and (36), an equation for determining the parameter $\gamma$ through $\Delta s_{tot}$ the signs "+" and "–" in $\Delta s_m$. This equation also has four solutions. These solutions are functions on $d = \Delta\langle s_{tot}\rangle/\langle K_0\rangle$ of the form:

$$\gamma_{01(+)(+)} = 4.38[\sqrt{1-0.67d}+1],$$
$$\gamma_{01(+)(-)} = 4.38[-\sqrt{1-0.67d}+1],$$
$$\gamma_{02(-)(+)} = 9.05[1+\sqrt{1+0.325d}],$$
$$\gamma_{02(-)(-)} = 9.05[1-\sqrt{1+0.325d}], \quad d = \Delta\langle s_{tot}\rangle/\bar{K}_0.$$
(65)

The notation "0" indicates that $\ln Z_\tau(s_{ef})_{\tau\to\infty} \to 0$. Figs. 5-8 show the results of calculations using expressions (29) and (65), corresponding to the results from Figs. 1-4 obtained using expressions (39) and (63). The behavior of the parameter $\gamma$ does not change qualitatively. As in the case of $\gamma_{1(+)(+)}$ (63), for case (65), the quantities $\gamma_{01(+)(+)}$ and $\gamma_{02(-)(+)}$ describe the system in a stationary non-equilibrium state, while the quantities $\gamma_{01(+)(-)}$ and $\gamma_{02(-)(-)}$ describe possible equilibrium states. The value $\gamma_{02(-)(-)}$ in (65) is realized only for small values of $d<0.0067$. However, for partition function (29), the convergence condition differs from (58). Therefore, the question of the existence of the values $\gamma_{02(-)(-)}$ and the boundaries of $\gamma_{02(-)(+)}$ requires separate study. Given the similarity of the results obtained using partition functions (39) and (29), it can be assumed that expression (65) may also be used to estimate the convergence of relation (29). The requirement for the radical expression in $\gamma_{01(+)(+)}$ and $\gamma_{01(+)(-)}$ to be positive provides the condition $d<1.49$.

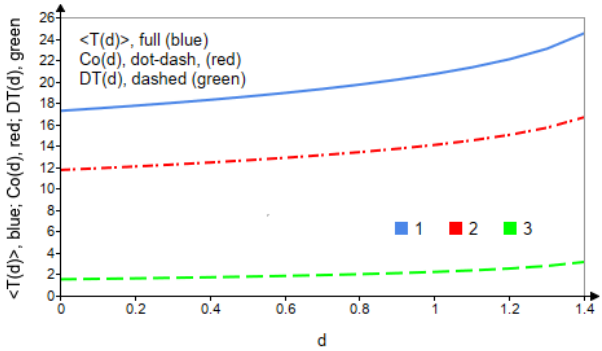

Fig. 5: Behavior of $<T(d)> = \langle T\rangle = \langle\tau_\gamma\rangle$ (sec) (60) [solid line, blue], $D_T(d)$ (sec$^2$) (61) [dashed line, green], and *CorrTK(d)* (sec) (62) [dot-dash, red], calculated using expressions (29) and (65) for $\gamma_{01(+)(+)}$ (65) in the range $0<d<1.4$.

It can be observed that the first and second moments of *FPT*, as well as the correlation between *FPT* and activity *K*, behave similarly to those in Fig. 1, but with a vertical shift.

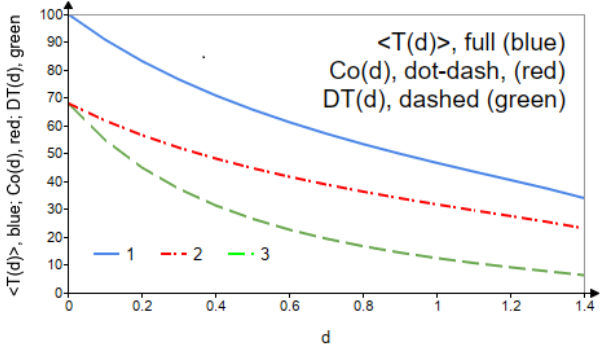

Fig. 6: Behavior of $<T(d)> = \langle T \rangle = \langle \tau_\gamma \rangle$ (sec) (60) [solid line, blue], $D_T(d)$ (sec²) (61) [dashed line, green], and *CorrTK(d)* (sec) (62) [dot-dash, red] on $d = \Delta \langle s_{tot} \rangle / \langle K_0 \rangle$, calculated using expressions (29) and (65) for $\gamma = \gamma_{01(+)(-)}$ (65) in the range $0 < d < 1.4$.

The behavior in Fig. 6 for partition function (28)-(29) mirrors that of Fig. 2 for partition function (39). The general nature of the dependencies remains unchanged.

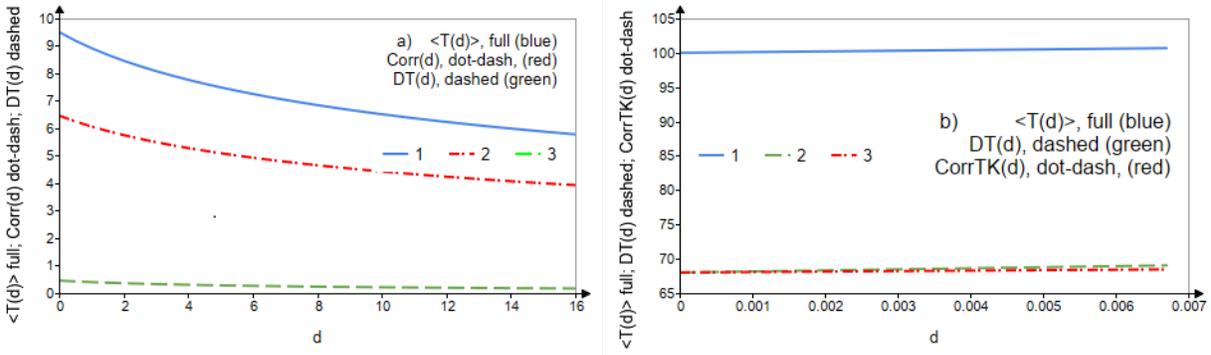

Fig. 7: Behavior of $<T(d)> = \langle T \rangle = \langle \tau_\gamma \rangle$ (sec) (60) [solid line, blue], $D_T(d)$ (61) [dashed line, green], and *CorrTK(d)* (62) [dot-dash, red], calculated using expressions (29) and (65) in two ranges: (a) for $\gamma = \gamma_{02(-)(+)}$ (65) in the range $0 < d < 15$ (Fig. 7a) and (b) for $\gamma = \gamma_{02(-)(-)}$ (65) in the range $0 < d < 0.007$ (Fig. 7b), with Fig. 7b corresponding to the convergence of the Laplace transform of the *FPT* distribution $Z_\gamma$, obtained from equation (58).

Fig. 7b aligns with the behavior in Fig. 3.

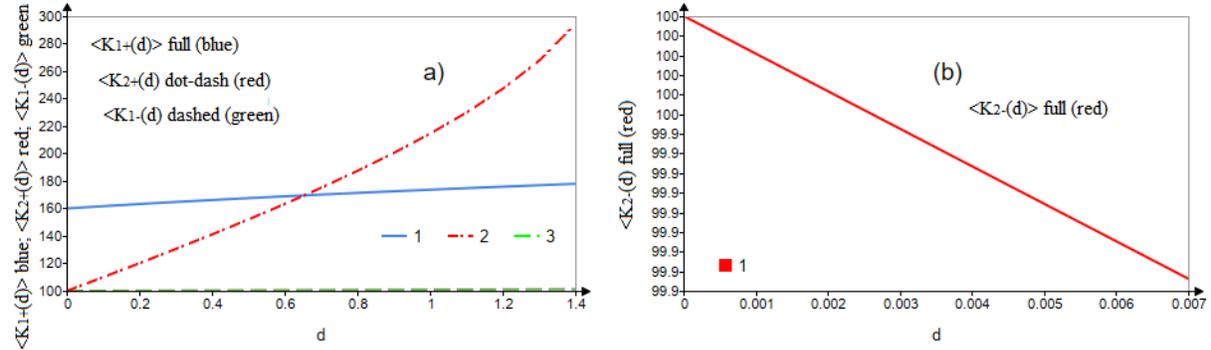

Fig. 8: Behavior of $\langle K_\gamma \rangle$ (64) as a function of $d = \Delta \langle s_{tot} \rangle / \langle K_0 \rangle$ depending on $\gamma_{01(+)(+)}$ (65), $\langle K_{\gamma 1+} \rangle$ [full (blue)], on $\gamma_{02(-)(+)}$ (65), $\gamma > 0$, $s < 0$ (active phase), $\langle K_{\gamma 2+} \rangle$ [dot-dash (red)], and on $\gamma_{01(+)(-)}$ (65), $\langle K_{\gamma 1-} \rangle$ [dashed (green)], $\gamma < 0$, $s > 0$ (inactive phase), calculated using expressions (29) and (65). Panel (a) shows $\langle K_{\gamma 1+} \rangle$, $\langle K_{\gamma 2+} \rangle$, and $\langle K_{\gamma 1-}\rangle$ in the range 0<d<1.4, while panel (b) shows $\langle K_{\gamma 2-}\rangle$ as a function of $\gamma_{02(-)(-)}$ in the range 0<d<0.007 (Fig. 8b), corresponding to the convergence of the Laplace transform of the *FPT* distribution $Z_\gamma$, obtained from equation (58).

Thus, the results obtained using expressions (29) and (39) are quite similar, even though the partition functions differ. Fewer assumptions were made when deriving expression (29), making it a potentially more preferable approach.

## 4C. Example: Classical Three-Level System

Three-level systems are present in various physical contexts, such as a spin-1 particle in an external magnetic field, the three states of atoms or molecules bound by lasers or other interactions, a three-level atom interacting with two classical monochromatic fields, and oscillations between three neutrino flavors. Here, we focus on the simplicity of stochastic Markovian classical systems, which are approximations of quantum systems under specific limits (e.g., large mass/energy, high temperature, long timescales). In [23], for a three-level system, the statistics of the number *K* of photons emitted from Level $|1\rangle$ decays to Level $|0\rangle$, with rate $\kappa_1$, are discussed. Let's consider a simple model, a special case of a three-level system, as explored in [32]. In this simplified model, we consider three configurations: $|0\rangle$, $|1\rangle$, and $|2\rangle$. We observe jumps between configurations $|2\rangle$ and $|0\rangle$, ignoring transitions to and from configuration $|1\rangle$. Using the notation from equations (22)-(25), we set *N*=1, and $\vec{M}$, the total number of transitions between the top and bottom levels is represented by $K_{20}$.

In the *s*-ensemble, the largest eigenvalue of the operator:
$$W_{s_{20}} = \kappa \begin{pmatrix} -1 & 1 & 0 \\ 0 & -1 & 1 \\ e^{-s_{20}} & 0 & -1 \end{pmatrix} \quad (66)$$

(where $s_{20}$ is the field conjugated to $K_{20}$, and $\kappa$ is the rate of jumps between configurations $|2\rangle$ and $|0\rangle$, associated with photon emission into the bath) gives the large deviation (*LD*) function:
$$\Theta(s_{s_{20}}) = \kappa(e^{-s_{20}/3} - 1). \quad (67)$$

This function is the cumulant generating function for the number of jumps $K_{20}$ per unit time. In the *x*-ensemble context, the relevant operator is:
$$T_{x,s_{20}} = \frac{\kappa}{x+\kappa}\begin{pmatrix} 0 & 1 & 0 \\ 0 & 0 & 1 \\ e^{-s_{20}} & 0 & 0 \end{pmatrix}. \quad (68)$$

From its largest eigenvalue, we obtain the *LD* function from equation (25):
$$G(\gamma, s_{20}) = -s_{20} + 3[\log\kappa - \log(\gamma + \kappa)]. \quad (69)$$

This is the generating function for the cumulants of both $\tau/K$ and $K_{20}/K$:
$$g(\gamma) = G(\gamma, s_{20} = 0) = 3[\log\kappa - \log(\gamma + \kappa)]. \quad (70)$$

Solving the equation $G(x_*, s_{20}) = 0$ for $x_*$, we obtain $x_*(s_{20}) = g_*(s_{20}) = \Theta(s_{20})$ from equation (67) and from equation (21).

Now, consider the case corresponding to expression (29), where at large times, the partition function equals unity. For the average value of $K$ from equations (10), (12), (41), (67), and (21), we obtain:

$$\langle K_\gamma \rangle = -\tau \frac{\partial \Theta(s)}{\partial s}\bigg|_{s\,=g(\gamma)} = $$
$$= \tau \frac{\kappa}{3} e^{-s_{20}/3}\bigg|_{s_{20}=g(\gamma)} = \tau \frac{\kappa}{3}(1+\frac{\gamma}{\kappa}). \quad (71)$$

This expression depends on $\tau$, which is not specified in equation (10); it only needs to be large enough to satisfy the *LD* relations. Thus, there is some arbitrariness in the choice of $\tau$. The average value of $\tau$ from equation (70) is:

$$\langle \tau_\gamma \rangle = -K \frac{\partial g(\gamma)}{\partial \gamma} = $$
$$= \frac{3K}{\kappa} \frac{1}{1+\gamma/\kappa} = \tau_0 \frac{1}{1+\gamma/\kappa}. \quad (72)$$

If we choose $K = \langle K_{\gamma=0} \rangle = K_0$ in equation (72) and substitute this value into equation (71), taking into account the relation $\gamma = x = x_*$, $e^{-s_{20}/3} = 1+\gamma/\kappa$, leads to the expression $\langle \tau_{\gamma=0} \rangle = 3K_0/\kappa$. If we substitute expression (72) into expression (71) with $\tau = \langle \tau_{\gamma=0} \rangle = 3K_0/\kappa$, we derive the following expression:

$$\langle K_\gamma \rangle = K_0(1+\gamma/\kappa). \quad (73)$$

Differentiating equations (72) and (71) with respect to $\gamma$ and $s$, respectively, we obtain the relations for the dispersions $D_\tau$ and $D_K$ of the time $\tau$ and the dynamic activity $K$:

$$D_\tau = \frac{3K_0}{\kappa^2} \frac{1}{(1+\gamma/\kappa)^2}, \quad (74)$$

$$D_K = \frac{K_0}{3}(1+\gamma/\kappa) = \frac{\langle K \rangle}{3}. \quad (75)$$

Using equations (34), (72), and (75), we find:

$$D_{\tau K} = \frac{K_0}{\kappa} \frac{1}{(1+\gamma/\kappa)}. \quad (76)$$

As in Sections 4A and 4B, it is possible to find the changes in entropy of the system $\Delta\langle s_{sys} \rangle$ using either expression (39) or expression (29). We use expression (29) as in Section 4B. Let us consider a simplified model where $\kappa = 0.5$, $K_0 = 100$. For $\Delta\langle s_{sys} \rangle$, assuming large times $\ln Z_\tau(s_{ef}) = 0$, we expand in a series in $\gamma$, restricting to quadratic terms: $\Delta\langle s_{sys} \rangle = 9K_0(\gamma/\kappa)^2/2$. Using expression (47), after expansion, we find: $\Delta s_m = \pm K_0 \gamma / \kappa$. Depending on the sign in front of $\Delta s_m$, we obtain two quadratic equations with the following solutions: with the "+" sign in $\Delta s_m$:

$$\gamma/\kappa = [\pm\sqrt{1+18d} - 1]/9, \quad d = \Delta s_{tot}/K_0, \quad (77)$$

and with the sign "-" in $\Delta s_m$:

$$\gamma/\kappa = [\pm\sqrt{1+18d} + 1]/9. \quad (78)$$

Condition (58) is not satisfied for solution (77) $(\gamma_{(+-)}/\kappa)(d)$ with the minus sign. For solutions $(\gamma_{(++)}/\kappa)(d) = [\sqrt{1+18d}-1]/9$, $(\gamma_{(-+)}/\kappa)(d) = [\sqrt{1+18d}+1]/9$, condition (58) is satisfied for all $d$. For $(\gamma_{(--)}/\kappa)(d) = [-\sqrt{1+18d}+1]/9$, condition (58) is satisfied for $d < 1/\tau_0 + 9/2\tau_0^2$. In [23], it is shown that for $s>0$ (which corresponds to $\gamma<0$), the cumulant $\theta(s)$ takes a constant value. However, we are considering a very small range of $\gamma<0$, where this can be neglected.

We rewrite the realized parameters in (77)-(78) in the form:

$$(\gamma_{(++)}/\kappa)(d) = [\sqrt{1+18d}-1]/9,$$
$$(\gamma_{(-+)}/\kappa)(d) = [\sqrt{1+18d}+1]/9, \qquad (79)$$
$$(\gamma_{(--)}/\kappa)(d) = [-\sqrt{1+18d}+1]/9.$$

The solution $(\gamma_{(-+)}/\kappa)(d)$ describes stationary nonequilibrium states, where no equilibrium is present. For solutions $(\gamma_{(++)}/\kappa)(d)$ and $(\gamma_{(--)}/\kappa)(d)$, equilibrium and phase transitions are possible.

Substituting expressions (79) into relations (72)-(76), we obtain the following figures:

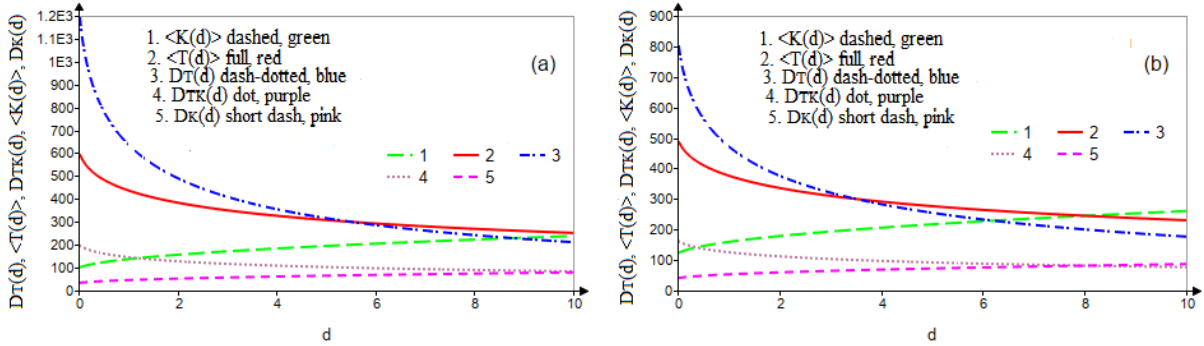

Fig. 9: Behavior of functions (72)-(76) $<T(d)>=\langle\tau_\gamma\rangle(d)$ [solid red, from equation (72)], $<K(d)>=\langle K_\gamma\rangle(d)$ [dashed green, from equation (73)], $D_T(d) = D_\tau(d)$ [dash-dotted blue, from equation (74)], $D_K(d)$) [short dash pink, from equation (75)], and $D_{TK}(d) = D_{\tau K}(d)$ [dotted purple, from equation (76)], depending on $d = \Delta\langle s_{tot}\rangle/\langle K_0\rangle$, when substituting parameter $\gamma(d)/\kappa$ (from equation (79)) into equations (72)-(76). The range is $d=(0,...,10)$ for $(\gamma_{(++)}(d)/k)$, Fig. 9a) and $(\gamma_{(-+)}(d)/k)$ (79) in Fig. 9b.

Figure 10 shows the dependence of average values and variances on $(\gamma_{(--)}(d)/k)$ (79) in the interval $d$ values $(0,...,0.0016)$. It can be seen that the average values and variances remain constant over this interval, confirming the conclusions from [23]. A similar situation is observed in Fig. 7b.

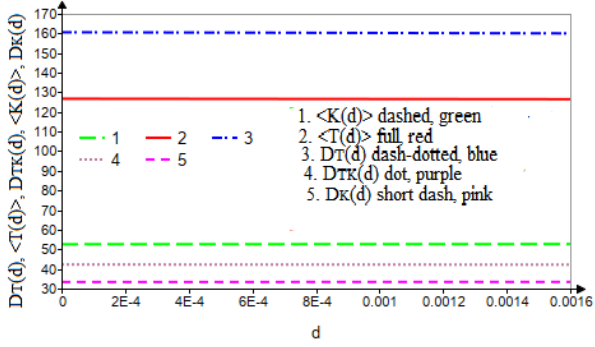

Fig. 10: Behavior of functions (72)-(76) $<T(d)> = \langle \tau_\gamma \rangle(d)$ [solid red, from equation (72)], $<K(d)> = \langle K_\gamma \rangle(d)$ [dashed green, from equation (73)], $D_T(d) = D_\tau(d)$ [dash-dotted blue, from equation (74)], $D_K(d)$ [short dash pink, from equation (75)], and $D_{TK}(d) = D_{\tau K}(d)$ [dotted purple, from equation (76)], depending on $d = \Delta \langle s_{tot} \rangle / \langle K_0 \rangle$, when substituting parameter $(\gamma_{(-)}(d)/k)$ (from equation (79)) into equations (72)-(76). The range is $d=(0,\ldots,0.0016)$.

## 4D. Example: The Quantum Two-Level System, Partition Function (80)

In subsections 4A)-4C), we discussed distributions (29) and (39). In this subsection, we introduce a different distribution of the form:

$$P_M^{sx}(K,\tau) = Z_M^{-1}(K,\tau) e^{-sK-x\tau} P_M(K,\tau), \quad (80)$$

where $M$ is a fixed value, as in (27). Here, $K$ and $\tau$ are fluctuating parameters, and $\tau$ represents the duration of an unbiased trajectory while $K$ represents the fluctuating observable. This distribution $P_M(K, \tau)$ is the probability that an unbiased trajectory has a duration $\tau$ and the observable $K$ reaches a certain value by the time the other observable reaches the fixed value $M$. This distribution was derived in works such as [88], [52-53], [83-84], and [32]. It is written based on expression (27), after summation over $M_1$.

The corresponding grand-partition function is:

$$Z_M(K,\tau) = \sum_K \sum_\tau e^{-sK-x\tau} P_M(K,\tau) \sim e^{MG(s,x)}, \quad (81)$$

which takes the expected large deviation form with a scaled cumulant generating function (*SCGF*) $G(s,x)$, as defined in (25). The connection of distributions of the form (80), (81) (as well as of the form (29), (39)) with stochastic dynamics and the Master Equation, Eqs. (1)-(2), is discussed in papers [9], [28], [32]. Paper [88] shows how the form of distributions and the partition function sums is connected with the random walk and the transition matrix.

We use the notation $G$, as in (25) [32], in [88] the notation $\varphi$ is used. The *SCGF* from (19) $g(x) = G(s,x)_{s=0}$ is used here. The average values are:

$$\begin{aligned}\langle K \rangle &= -\tau_0 \partial g(\gamma)/\partial \gamma, \\ \langle \tau \rangle &= -K_0 \partial \theta(s)/\partial s.\end{aligned} \quad (82)$$

Let's apply these expressions to the quantum two-level system.

The quantum two-level system represents a scenario in which two quantum levels $|0\rangle$ and $|1\rangle$ are coherently driven on resonance at a Rabi frequency $\Omega$ and coupled to a zero-temperature bath [49]. The Hamiltonian, single-jump operator, super-operator $W_s$, the operator $T_{x,s}$ (26), and its largest eigenvalues are presented in [32]. The largest eigenvalues $G(s,\gamma), g(\gamma)$ correspond to those in expressions (67), (69), and (70) of Section 4C), with only the parameter value differing. Accordingly, expressions (72)-(76), (77)-(79) also coincide.

In this case, the parameter $\kappa$ from Section 4C) is replaced by $2\Omega$, with $\Omega=1$ MHz:

$$\langle \tau_{\gamma_1} \rangle = 150/(1+\gamma_1),$$
$$\langle K_{\gamma_1} \rangle = 100(1+\gamma_1),$$
$$\langle D_\tau \rangle = 75/(1+\gamma_1)^2, \quad (83)$$
$$\langle D_K \rangle = \langle K_{\gamma_1} \rangle /3,$$
$$\langle D_{\tau K} \rangle = \langle \tau_{\gamma_1} \rangle /3,$$
$$\gamma_1 = \gamma/2\Omega = \gamma/2.$$

The largest eigenvalues yield the *LD* function from equations (10), (19), and (25). We set $K_0 = M = 100$:

$$G(s,x) = -s - 3\ln(1+\gamma/2\Omega),$$
$$\theta(s) = 2\Omega(e^{-s/3} - 1),$$
$$g(\gamma) = -3\ln(1+\gamma/2\Omega).$$

Entropy of the system $\langle s_{sys} \rangle$ and $\Delta \langle s_m \rangle$, is given by:

$$\langle s_{sys} \rangle / M = s\langle K \rangle + \gamma\langle \tau \rangle + G(s,\gamma),$$
$$\Delta \langle s_{sys} \rangle = -\langle s_{sys} \rangle,$$
$$\Delta \langle s_m \rangle = -\tau_0 [\partial \theta(s)/\partial s - \partial \theta(s)/\partial s |_{s=0}].$$

Using equation $s = g(\gamma)$ (21), $G(s,x) = 0$, and $\ln Z_M(K,\tau) = 0$ as in (29), we can write the behavior of the random variables $\tau$ and $K$ similarly to expression (29). The behavior of the average and second moments of $\tau$ and $K$ are shown in Fig. 11:

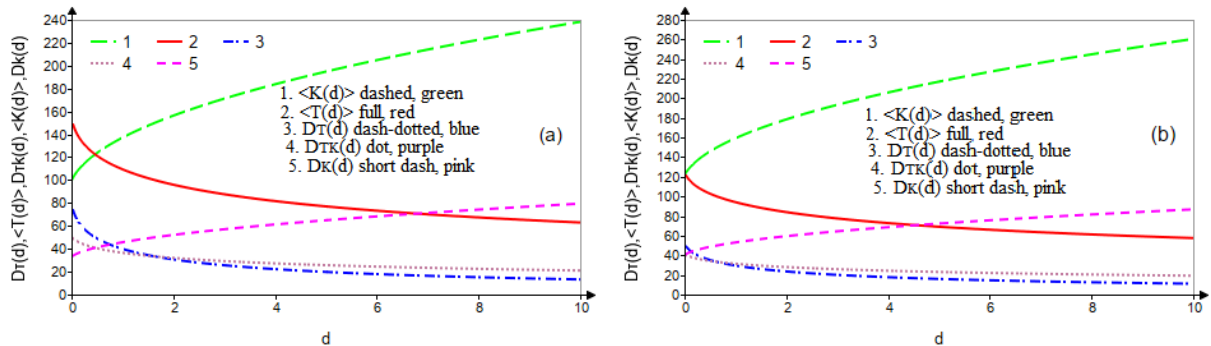

Fig. 11: Behavior of the functions (83) $<T(d)>= \langle \tau_{\gamma_1} \rangle$ [solid red], $<K(d)>= \langle K_{\gamma_1} \rangle$ [dashed green], $D_T(d) = D_\tau(d)$ [dash-dotted blue], $D_K(d)$ [short dashed pink], and $D_{TK}(d) = D_{\tau K}(d)$ [dotted purple], depending on $d = \Delta\langle s_{tot} \rangle / \langle K_0 \rangle$. The parameter $\gamma(d)/2\Omega$ from equation (79) is substituted into expressions (83). The range is $d=(0,\ldots,10)$, $(\gamma_{(++)}(d)/2\Omega)$ in Fig. 11a), and $(\gamma_{(-+)}(d)/2\Omega)$ in Fig. 11b).

The solution for $(\gamma_{(+-)}/\kappa)(d)$ is not realized in this case, similar to Section 4C. The negative branch of the solution for $(\gamma_{(--)}/\kappa)(d)$ is constant in the interval $d=(0,\ldots,0.0067)$, as seen in Fig. 10.

In Appendix B, the parameter $\gamma$ is related to other physical parameters of the system, and the physical meaning of the parameter $s$ (and the associated parameter $\gamma$) is discussed in further detail.

## 5. Conclusion

In [83-84], first passage time (*FPT*) was treated as an independent thermodynamic variable in the statistical ensemble. A similar concept is explored in the thermodynamics of trajectories, where trajectories with random time (*FPT*) are studied [33], and the partition function is determined using large deviation (*LD*) theory. This paper builds upon those approaches by investigating the dependence of average *FPT*, *FPT* dispersion, dynamic activity, and the correlation between *FPT* and dynamic activity on total entropy change. The entropy change comprises both the intrasystem entropy production and the entropy exchange with the environment.

In [84], the Gibbs canonical ensemble was generalized to nonequilibrium by introducing *FPT* as an additional thermodynamic parameter. A similar procedure is carried out in the thermodynamics of trajectories for *x*-ensembles, as shown in [32, 33]. This paper applies the approach of [84] to the formalism of trajectory thermodynamics, examining the joint distribution of dynamic activity *K* (which can be generalized to other observables) and *FPT*. While [84] assumed the independence of energy and *FPT*, this work explores the dependence between random distribution parameters. Using *LD* theory, we derived the partition function for large dynamic activity *K* and relatively small *FPT* values, expressed as an *LD* function in equation (29). The correlation between *K* and *FPT*, along with their first and second moments, was obtained analytically.

The importance of *FPT*, introduced in the beginning, is crucial for understanding its behavior, particularly its dependence on entropy change. This study provides a detailed analysis of physical effects associated with entropy changes in a system, calculating three model systems using different distributions. Despite the different behaviors of moments in the examples, three branches of dependence between moments and entropy change were realized in all cases (except for 4B)). Two branches describe active states where $\gamma>0$ (and $s<0$). In one of these branches, equilibrium states and phase transitions are possible, while the other describes stationary nonequilibrium states without equilibrium or phase transition. The solution with $\gamma<0$ corresponds to near-equilibrium states and exists only for minimal entropy changes. The results in Section 4 confirm the theoretical conclusions from [23].

A key distinction between this paper and [32] is the focus on averaging over fixed values of *K*, assuming an ensemble of systems with fixed *K*. This approach allows us to find the correlation between *K* and $\tau$, which intuitively correlates with *FPT*: the longer the *FPT*, the more events occur.

We obtained the analytical form of the dependences of average dynamic activity, *FPT*, their variances, and their correlation on the parameter conjugate with *FPT*. A major contribution is linking this parameter with total system entropy change, thereby expressing system behavior through entropy. Knowing these dependencies provides avenues for optimal system control, for instance, by leveraging the relationship between *FPT* and total entropy change.

The *FPT* moment dependences on entropy change were calculated for three models using expressions (29), (39), and (80). Close behaviors were observed for the partition functions, though some conditions, such as the convergence of (29), had to be modified for practical use. We also explored the dependence of $\gamma$ on entropy change in nonequilibrium states, confirming that only positive γ is realized, with no phase transitions.

Throughout this work, we noted the analogy between trajectory thermodynamics and the nonequilibrium statistical operator (*NSO*) method, where both theories consider the system's history but use different approaches to time averaging. In trajectory thermodynamics, large deviation theory plays a key role, while the *NSO* method relies on the maximum entropy principle.

The results are valid for simple two-level, three-level, and quantum two-level systems. For other systems, the outcomes might vary depending on the function $g(\gamma)$. However, certain general results, such as the restriction of negative $\gamma$ by the convergence of the Laplace transform for the FPT distribution, remain valid.

There are still open questions, particularly regarding which approximation best describes specific physical situations. For example, what physical insights are gained from distributions (29) versus (39)? The findings in this paper are summarized as follows:

**1). Analysis of Two-Level and Three-Level Systems**: Using classical and quantum two-level and three-level systems as examples, we demonstrated how these states can be analyzed by considering the behavior of the parameter $\gamma$ (or *x*) conjugate to random time $\tau$. This enabled us to differentiate between stationary nonequilibrium states and equilibrium states.

2). **Dependence of Moments on Entropy Change**: The dependences of moments (mean values, variances, correlations) of random variables like dynamic activity and observation time on total entropy change $\Delta s_{tot}$ were derived. These results can be used to evaluate how moments change with $\Delta s_{tot}$, which includes entropy exchange with the environment $\Delta s_m$. This entropy is crucial in trajectory thermodynamics (e.g., [42]-[45]). The example of electron transfer in the Coulomb blockade regime [106] demonstrated the significant influence of entropy change on transfer speed as the number of electrons reaches a critical value.

**3). Connection to Nonequilibrium Thermodynamics**: This approach helps to connect trajectory thermodynamics with nonequilibrium thermodynamics, particularly stochastic thermodynamics. The relationship between expressions (42)-(45) and thermodynamic forces and flows (e.g., [101-102], [113]) is explored. Similar topics are studied in [112], focusing on the impact of heat and work on system behavior. Knowing the influence of $\Delta s_m$ on the behavior of the system, we can consider the impact of thermodynamic forces and flows on this behavior.

**4). Analogies with Classical Thermodynamics**: By drawing analogies between trajectory thermodynamics and classical thermodynamic distributions, we translated methods for obtaining stationary thermodynamic characteristics into the context of trajectory dynamics. This highlights the potential of *LD* approaches in predicting system behavior based on entropy change.

Overall, this paper establishes a foundational link between entropy changes and *FPT* within the framework of trajectory thermodynamics, offering insights into how non-equilibrium entropy influences dynamic activity and *FPT*. These findings advance theoretical understanding and open future research directions for thermodynamic models in complex systems.

# Appendix A. Non-zero values of the parameter γ for the values of K and τ

Let us expand the value $s \neq 0$, $\langle K \rangle = \langle K_s \rangle = \langle K_\gamma \rangle = -\dfrac{\partial \theta(s)}{\partial s}\bigg|_{s=g(\gamma)} \tau$ from equation (35) in a series around $\gamma$, limiting ourselves to the linear term. As we will see, this approximation is sufficient to determine the value of $\gamma$ from subsequent expressions, while terms involving $\gamma^2$ refine the expansion further. To proceed, consider the partition function in equation (10), which includes the probability distributions from equations (7)-(8). We expand this probability $P_{\langle \tau_\gamma \rangle}$, $\tau = \langle \tau_\gamma \rangle$ in a series around $\gamma$, taking into account the expansion obtained from equations (9)-(10). Restricting ourselves to the first-order term in $\gamma$, we have: $P_\tau(K) = e^{-\tau\varphi(K/\tau)}$,

$P_{\langle \tau_\gamma \rangle} \cong P_{\langle \tau_{\gamma=0} \rangle} + \gamma(\partial P_{\langle \tau_\gamma \rangle} / \partial \langle \tau_\gamma \rangle)(\partial \langle \tau_\gamma \rangle / \partial \gamma)|_{\gamma=0}$, $\partial P_{\langle \tau_\gamma \rangle} / \partial \langle \tau_\gamma \rangle = P_{\langle \tau_\gamma \rangle}[-\varphi(K/\tau) + (K/\tau)(\partial \varphi(K/\tau)/\partial(K/\tau))]$, $\tau = \langle \tau_\gamma \rangle$.

We believe $\tau \neq \tau_0$, $\tau = \langle \tau_\gamma \rangle$.

For the two-level quantum system driven coherently at a Rabi frequency $\Omega$, the explicit form of the function $\varphi$ is taken from [5], where: $\varphi(\kappa) = 3[\kappa \ln(\kappa/b) - (\kappa - b)]$, $b = 2\Omega/3 = \langle K \rangle/\tau$.

The calculation performed gives the following result:

$$\langle K_\gamma \rangle = \langle K_0 \rangle - \gamma B_k, \tag{A.1}$$

where $B_k = 6c_3 D_{K_0}/a_1$, $c_3 = a_1^2/2 - a_2 = \dfrac{\kappa^2 + \eta^2}{2(\kappa\eta)^2}$. Taking into account that $\langle K_\gamma \rangle$ depends on $\gamma$, from expressions (19), (A.1), we obtain by $K = \langle K_\gamma \rangle$, $\ln Z_K(\gamma) = \langle K_\gamma \rangle g(\gamma)$:

$$\begin{aligned}\langle \tau_\gamma \rangle &= -\dfrac{\partial \ln Z_K(\gamma)}{\partial \gamma} \\ &= -\dfrac{\partial \langle K_\gamma \rangle}{\partial \gamma} g(\gamma) - \langle K_\gamma \rangle \dfrac{\partial g(\gamma)}{\partial \gamma}.\end{aligned} \tag{A.2}$$

For $\Delta\langle s_m \rangle$ and $\langle\tau_\gamma\rangle$ we get $\Delta\langle s_m \rangle = [\langle K_0 \rangle \partial g(\gamma)/\partial \gamma|_{\gamma=0} - (\langle K_0 \rangle - \gamma B_k)\partial g(\gamma)/\partial \gamma + B_k g(\gamma)]\partial \theta(\lambda)/\partial \lambda|_{\lambda=0}$, and

$$\langle \tau_\gamma \rangle = -\langle K_\gamma \rangle \partial g(\gamma)/\partial \gamma = -[\Delta\langle s_m \rangle/\Delta s_1 + K_0]\partial g(\gamma)/\partial \gamma.$$

We obtain from (21), (31), (39)-(41):

$$\langle s_\gamma \rangle = \langle s_{sys} \rangle = -\langle \ln p_{s\gamma} \rangle, \tag{A.3}$$

where $\ln Z_s = \langle \tau_\gamma \rangle \theta(s)$, $\ln Z_\gamma = \langle K_\gamma \rangle g(\gamma)$. We will evaluate for $\tau_0 \to \langle \tau_\gamma \rangle$, $K_0 \to \langle K_\gamma \rangle$.

Since at $s = 0$, $\theta(s) = 0$, $\gamma = 0$, $g(\gamma) = 0$, then:

$$\langle s_0 \rangle = \langle s_{sys|\gamma=0} \rangle = s\langle K_0 \rangle + \ln Z_{s|\gamma=0} = 0, \tag{A.4}$$

$$\begin{aligned}\Delta\langle s_{sys} \rangle &= \langle s_0 \rangle - \langle s_\gamma \rangle = -s\langle K_\gamma \rangle - \gamma\langle \tau_\gamma \rangle - \ln Z_\gamma - \ln Z_s \\ &= -2g(\gamma)\langle K_\gamma \rangle - 2\gamma\langle \tau_\gamma \rangle.\end{aligned} \tag{A.5}$$

Using the relation obtained in [96] from the general expression (45) of the theory of large deviations, $\langle s_m \rangle = -\langle \tau \rangle \partial \theta(\lambda)/\partial \lambda$, assuming $\lambda=0$ (although arbitrary $\lambda$ and the biased ensembles of trajectories are possible in [96]), from (47) (A1) we obtain:

$$\Delta \langle s_m \rangle = [-\langle K_0 \rangle a_1 + (\langle K_0 \rangle - \gamma B_k) \frac{a_1 + 2\gamma a_2}{1 + \gamma a_1 + \gamma^2 a_2} -$$
$$-B_k \ln(1 + \gamma a_1 + \gamma^2 a_2)] \frac{\partial \theta(\lambda)}{\partial \lambda}\bigg|_{\lambda=0}.$$
(A.6)

## Appendix B. Connection of Parameter γ with Other System Parameters

In previous sections, the parameter $\gamma$ was primarily associated with changes in the system's entropy. This included entropy exchanges with the environment, as discussed, for example, in Section 4. While the connection between $\gamma$ and entropy changes was established, there are other possible interpretations and calculations for determining $\gamma$.

In references [52-53, 84], a distribution involving random variables for energy $u$ and first passage time (*FPT*) $T_\gamma$ is described. This distribution resembles equation (39) when the substitution $s \to \beta$, $K \to u$, $\tau \to T_\gamma$ is made, where $\beta$ is the inverse temperature and $u$ the internal energy density [52, 83-84]. It's important to note that these distributions differ even though the parameter $\gamma$ is conjugate to the random variable $T_\gamma$ in both cases. The additional variables $\beta u$ are different from the pair $sK$ in equation (39), creating an analogy rather than an exact equivalence. Equating these distributions essentially implies $s = \beta$, but this is more of an interpretative analogy.

For the distribution with random energy and time parameters, relations (21) that connect $\gamma$ and $s$, which hold for equation (39), do not directly apply. However, an analogous interpretation is possible. Below, we present the physical meaning of the parameter $\gamma$ for a distribution other than (39). Note that this discussion is based on an analogy with the parameter $\gamma$ from the thermodynamics of trajectories and not an exact match.

Starting with a distribution containing random variables for energy $u$ and *FPT* $\tau = T_\gamma$ [52, 83-84], we derive the internal entropy $s_\gamma$ and its differential:

$$s_\gamma = -\langle \ln \rho(z; u, T_\gamma) \rangle = \beta \langle u \rangle + \gamma \langle T_\gamma \rangle + \ln Z(\beta, \gamma);$$
$$ds_\gamma = \beta d\langle u \rangle + \gamma d\langle T_\gamma \rangle,$$
(B.1)

where $\gamma$ is the conjugate variable for time.

From (B.1), we obtain the relation ($T$ is temperature, $\beta \sim 1/T$):

$$\gamma = \frac{\partial s_\gamma}{\partial \langle T_\gamma \rangle}\bigg|_{\langle u \rangle} = -\frac{\partial \langle u \rangle}{\partial \langle T_\gamma \rangle}\bigg|_{s_\gamma} (\frac{\partial \langle u \rangle}{\partial s_\gamma}\bigg|_{\langle T_\gamma \rangle})^{-1} =$$
$$-\frac{1}{T} \frac{\partial \langle u \rangle}{\partial \langle T_\gamma \rangle}\bigg|_{s_\gamma}, \quad \beta = \frac{\partial s_\gamma}{\partial \langle u \rangle}\bigg|_{\langle T_\gamma \rangle}.$$
(B.2)

Since $\partial \langle T_\gamma \rangle = -\chi_{t \in T_\gamma} \partial t$, $\chi_{t \in T_\gamma} = \begin{cases} 1, t \in T_\gamma \\ 0, t \notin T_\gamma \end{cases}$, then from (B2) we obtain that:

$$\gamma = \frac{1}{T} \frac{\partial \langle u \rangle}{\partial t}\bigg|_{s_\gamma} \chi_{t \in T_\gamma}.$$

The expressions for $\gamma$ and $\beta$ are symmetrical. It is possible to further transform expression (B.2).

Let us carry out the same operations for a distribution of the form (39), [90]:

$$p_{s\gamma}(K,\tau) = e^{-sK-\gamma\tau} / Z_{s\gamma},$$
$$Z_{s\gamma} = \sum_K \int d\tau e^{-sK-\gamma\tau} \omega(K,\tau). \tag{B.3}$$

where, as in [85-86], the probability $P(K,t)$ is denoted by $\omega(K,t)$. The Gibbs-Shannon entropy is written similarly to (B.1) as:

$$s_\gamma = -\langle \ln p_{s\gamma}(K,\tau) \rangle = s\langle K \rangle + \gamma\langle \tau \rangle + \ln Z_{s\gamma},$$
$$ds_\gamma = sd\langle K \rangle + \gamma d\langle \tau \rangle. \tag{B.4}$$

From (B.4) we obtain:

$$\gamma = \frac{\partial s_\gamma}{\partial \langle \tau \rangle}\bigg|_{\langle K \rangle} = -\frac{\partial \langle K \rangle}{\partial \langle \tau \rangle}\bigg|_{s_\gamma} \left(\frac{\partial \langle K \rangle}{\partial s_\gamma}\right)^{-1}\bigg|_{\langle \tau \rangle} = -s\frac{\partial \langle K \rangle}{\partial \langle \tau \rangle}\bigg|_{s_\gamma}. \tag{B.5}$$

We write the quantity $\frac{\partial \langle K \rangle}{\partial \langle \tau \rangle}\bigg|_{s_\gamma}$ from (B.5) in the form $\frac{\partial \langle K \rangle}{\partial \langle \tau \rangle}\bigg|_{s_\gamma} = -\frac{\partial s_\gamma}{\partial \langle \tau \rangle}\bigg|_{\langle K \rangle} / \frac{\partial s_\gamma}{\partial \langle K \rangle}\bigg|_{\langle \tau \rangle}$.

Substituting these quantities into (B.5), we obtain from (B4):

$$\gamma = -s\frac{\partial \langle K \rangle}{\partial \langle \tau \rangle}\bigg|_{s_\gamma} = s\frac{\gamma - \langle K \rangle / D_{K\tau}}{s - \langle K \rangle / D_K}. \tag{B.6}$$

where $D_K$ is the dispersion of $K$ (32), $D_{K\tau}$ is a correlation between the parameters $K$ and $\tau$ (34). This shows a symmetry between $\gamma$ and $\beta$, with both playing analogous roles in relation to time and energy, respectively. This interpretation of $\gamma$ can be further transformed depending on the system under study.

Using large deviation theory (LD), for a two-level system, we can substitute expressions (60)-(62) into equation (B6) and derive a relation for $\gamma$, expressed in terms of the parameters $\eta$ and $\kappa$. Additionally, the parameter $s$ can be expressed through $\gamma$ via equation (21), and $K$ is approximated by its mean value $K_0$ as given in equation (30).

There are other interpretations of the physical meaning of the parameter $\gamma$. In some cases, the parameter $\gamma$ is directly related to the parameter $s$, which governs the system's counting field (equation (21)). While the $s$-field is not always physically tunable, its influence on the system's statistical behavior can significantly affect the shape of the overall distribution. In large deviation theory, the function associated with $s$ behaves like a free energy function, where $s$ acts similarly to an inverse temperature. In this context, $K$ is the time-extensive order parameter (the number of transitions), and $s$ is the conjugate "counting" field. The activity $K$ and counting field $s$ serves as extensive and intensive variables, respectively. Critical features of this parameter's physical interpretation — such as the critical values of $s$, zeros of moment-generating functions, etc. — have been explored in [107-115].

The physical interpretation of parameters $s$ and $\gamma$ remains system-dependent and context-specific. It varies according to the physical model, system parameters, and the specific phenomena under investigation. While the general relationship between $s$, $\gamma$, and entropy provides insights, there is no universal interpretation of these parameters applicable to all systems and models. Thus, each case must be analyzed individually, with the physical context dictating the appropriate interpretation of $\gamma$ and $s$.